\documentclass[acmlarge,screen,nonacm]{acmart}
\usepackage{multirow}
\usepackage{booktabs}
\usepackage{longtable}
\usepackage{multicol}
\usepackage[table,xcdraw]{xcolor}
\usepackage{float}
\usepackage{caption}
\usepackage{booktabs}
\usepackage{tabularx}
\usepackage{array}

\usepackage{ragged2e}
\usepackage{fvextra}
\DefineVerbatimEnvironment{Prompt}{Verbatim}{breaklines=true}

\usepackage{xcolor}
\usepackage[normalem]{ulem}

\newif\ifshowchanges
\showchangesfalse

\ifshowchanges
  \newcommand{\revise}[1]{\textcolor{red}{#1}}
  \newcommand{\delete}[1]{\textcolor{red}{\sout{#1}}}
\else
  \newcommand{\revise}[1]{#1}
  \newcommand{\delete}[1]{}
\fi

\AtBeginDocument{%
  }

\setcopyright{none}

\renewcommand\footnotetextcopyrightpermission[1]{}

\begin{document}



\title{Measuring Smartphone User Experience through a Hierarchical Metric Framework via Social Media Reviews}



\author{Xiaoteng Pan}
\affiliation{%
  \institution{Nankai University}
  \city{Tianjin}
  \state{Tianjin}
  \country{China}
}
\email{2120240803@mail.nankai.edu.cn}

\author{Mingang Lan}
\affiliation{%
  \institution{Nankai University}
  \city{Tianjin}
  \state{Tianjin}
  \country{China}}
\email{2120250756@mail.nankai.edu.cn}

\author{Chenrui Zhang}
\affiliation{%
  \institution{Nankai University}
  \city{Tianjin}
  \state{Tianjin}
  \country{China}}
\email{2120230804@mail.nankai.edu.cn}

\author{Yu Su}
\affiliation{%
  \institution{Nankai University}
  \city{Tianjin}
  \state{Tianjin}
  \country{China}}
\email{2120230794@mail.nankai.edu.cn}

\author{Weijie Liu}
\affiliation{%
  \institution{Nankai University}
  \city{Tianjin}
  \state{Tianjin}
  \country{China}}
\email{2120230783@mail.nankai.edu.cn}

\author{Yue Gao}
\affiliation{%
  \institution{Nankai University}
  \city{Tianjin}
  \state{Tianjin}
  \country{China}}
\email{2120230814@mail.nankai.edu.cn}

\author{Nan Gao}
\affiliation{%
  \institution{Nankai University}
  \city{Tianjin}
  \state{Tianjin}
  \country{China}}
\email{nan.gao@nankai.edu.cn}

\author{Haining Zhang}
\authornote{Corresponding author.}
\affiliation{%
  \institution{Nankai University}
  \city{Tianjin}
  \state{Tianjin}
  \country{China}}
\email{zhanghaining@nankai.edu.cn}

\renewcommand{\shortauthors}{Xiaoteng et al.}

\begin{abstract}
Smartphone user experience (UX) is widely expressed in user-generated online discourse across platforms, creating opportunities for in-the-wild measurement at scale. However, existing UX instruments and review-mining approaches do not provide a smartphone-oriented, theory-grounded hierarchical measurement specification that supports consistent aggregation and comparison across heterogeneous platforms. In this research, we propose a hierarchical smartphone UX measurement framework and an interpretable computational pipeline that translates cross-platform reviews into structured UX metrics. The pipeline extracts localized experience evidence units, maps them to the hierarchy via coarse-to-fine classification, and quantifies evaluations with a unified five-level satisfaction sentiment model. We apply the approach to a stratified subset of approximately 20,000 Chinese social media reviews covering four major smartphone brands across three platforms. The resulting metrics separate what users discuss, captured by normalized mention frequency, from how they evaluate it, captured by mean five-level sentiment scores. This work provides a scalable and interpretable evidence base for cross-brand comparison and metric-level interpretation beyond raw review volume or single-platform observations.

\end{abstract}




\maketitle

\section{Introduction}
\revise{Smartphones are among the most pervasive personal computing platforms}
through which people access digital services, communication, work, entertainment, and everyday routines at a global scale~\cite{joshhowarthHowManyPeople2021}. 
As of April 2026, there were about 5.83 billion unique mobile users worldwide, and smartphones accounted for most active handsets~\cite{datareportalDigitalWorld2026}. Global smartphone shipments reached 1.26 billion units in 2025~\cite{idcSmartphoneMarketShare2026}, highlighting the market's global scale and competitive intensity.
At this scale, even small differences in day-to-day experience can affect how users evaluate devices and compare brands~\cite{malaquiasSmartphoneUsersSatisfaction2020,lucapetruzzellisMobilePhoneChoice2010}, making smartphone UX consequential for mobile and ubiquitous computing research as well as product practice~\cite{akhilmathurMovingMarketResearch2017}.

Smartphone UX is not limited to whether users can complete specific tasks~\cite{maxhortSurveyPerformanceOptimization2022}. It is also shaped by how devices and systems behave in real use~\cite{liuguohaPerceivedImpactProduct2025,weifenghuResearchBrandImage2023}, including perceived responsiveness and stability, battery endurance, and the quality of visual and audio output~\cite{jiegaoExploringSmartphoneUser2024,chengjunzhouPaperResearchMobile2024}. Beyond functional and sensory qualities, smartphones also support self-expression~\cite{jielouEffectsMobileIdentity2022,russellw.belkExtendedSelfDigital2013} and social connectedness~\cite{sarahdiefenbachSmartphonePacifierIts2019} by enabling symbolic identity expression and \revise{interpersonal connection, and a sense of belonging.}
These characteristics make smartphone UX a broad and layered measurement object rather than a single usability or satisfaction outcome. 
They also make smartphones a frequent subject of public discussion, as users increasingly describe, compare, and evaluate device experiences through social media posts, product reviews, and platform-specific online communities.

Research in HCI and related fields has proposed many instruments and analytic approaches for assessing user experience, ranging from standardized questionnaires and lab-based protocols to log-based behavioral measures. These approaches have provided important ways to capture users' perceptions and evaluations. However, when the goal is to characterize smartphone UX at scale from in-the-wild user discourse, two limitations remain. First, existing UX scales and broad experience indicators are not organized as a smartphone-oriented construct space that can consistently map diverse review expressions to comparable UX metrics. As a result, analyses often remain fragmented \cite{lilishiProductFeatureExtraction2023}. Different research traditions focus on different parts of smartphone experience, such as usability, visual aesthetics, or sentiment \cite{youngryuDecisionModelsComparative2007,sebastiana.c.perrigSmartphoneAppAesthetics2023,yeyiranAspectbasedSentimentAnalysis2019}, leaving other facets underspecified or unmeasured. This partial coverage also makes findings difficult to interpret and compare across brands and platforms, because the construct boundaries and label meanings are not consistently defined. Second, prior review-based UX studies frequently rely on post-hoc topic discovery or platform-specific categorizations \cite{chengyangOnlineUserReview2021,yeyiranAspectbasedSentimentAnalysis2019}. These methods can reveal salient themes, but they do not offer a stable hierarchy for aggregating evidence or supporting controlled comparisons. The central problem addressed in this paper is therefore not merely how to classify smartphone reviews, but how to operationalize smartphone UX as a structured, evidence-traceable, and comparable measurement object from naturally occurring user discourse.

\revise{
This study examines naturally occurring smartphone reviews as in-the-wild expressions of user experience and investigates how a hierarchical measurement framework can support cross-brand and cross-platform characterization of expressed smartphone UX.
To address this goal, we propose an integrated review-based smartphone UX measurement approach that combines a hierarchical UX framework with an interpretable computational pipeline.
}
The framework was developed through literature synthesis, smartphone-context adaptation, review-expression examination, and expert review. 
It organizes smartphone UX into Functional, Sensory, and Social Experience, with lower-level metrics and explicit construct boundaries. 
The pipeline transforms raw reviews into localized experience evidence units, maps these units to the hierarchy through coarse-to-fine classification, and assigns each unit a unified five-level satisfaction score.
We apply this framework and pipeline to Chinese social media reviews of flagship smartphones from four major brands across three platforms. 
Using a stratified subset of approximately 20,000 reviews, we report metric-level salience and valence, distinguishing what users discuss from how they evaluate it. 
We further examine platform-stratified patterns within each brand to show how review-based UX measurements vary across platform contexts.

Building on this study, we make three contributions:

\begin{enumerate}
  \item We develop a theory-grounded hierarchical framework that defines smartphone UX as a structured measurement space. It clarifies the roles and boundaries of Functional, Sensory, and Social Experience, and provides lower-level metrics \revise{that can be operationalized} for review-based UX measurement.

  \item We implement an evidence-traceable computational pipeline for operationalizing the framework on cross-platform user reviews. By linking raw text, localized evidence units, metric labels, and satisfaction scores, the pipeline enables controlled aggregation and comparable reporting across brands and platforms.

  \item We provide a cross-brand and cross-platform empirical characterization of smartphone UX as expressed experience. By separating salience from valence, the analysis shows how review-based UX measurement can identify visible strengths, friction points, and platform-shaped expression patterns without treating online reviews as population-average satisfaction.
\end{enumerate}

\section{Related Works}
This section situates our work within prior research on smartphone UX measurement, review-based experience analysis, and computational opinion mining. We organize the review into three parts, corresponding to measurement frameworks, review-based UX studies, and text-based analysis methods.
\revise{We first review how UX and mobile experience frameworks define the construct space of smartphone UX. We then discuss online reviews as situated, post-purchase expressions of everyday smartphone experience. Finally, we examine computational review analysis methods and identify the need for a construct-grounded measurement schema that can connect unstructured review expressions to interpretable UX metrics.
}

\subsection{Smartphone UX Measurement and Construct Frameworks}
\label{2.1}

User experience (UX) is defined as \textit{``user's perceptions and responses that result from the use and/or anticipated use of a system, product, or service''} \cite{internationalorganizationforstandardizationErgonomicsHumansystemInteraction2019}. 
\revise{This definition already places UX beyond narrow task-performance usability and frames experience as users' responses to products, systems, and services.}
\revise{Prior UX research similarly describes experience} as holistic, combining cognitive, affective, and sensory responses~\cite{marchassenzahlExperienceDesignTechnology2010,jodiforlizziUnderstandingExperienceInteractive2004}, and spanning pragmatic qualities (usefulness and efficiency) as well as hedonic and symbolic qualities (stimulation, aesthetics, identification, and self-expression)~\cite{marchassenzahlThingUnderstandingRelationship2003,johnmccarthyTechnologyExperience2004,russellw.belkExtendedSelfDigital2013,marchassenzahlNeedsAffectInteractive2010}. 
\revise{Accordingly, UX evaluation should consider not only usefulness, efficiency, and ease of operation, but also affective reactions, aesthetics, identification, self-expression, and socially situated meanings \cite{internationalorganizationforstandardizationErgonomicRequirementsOffice1998,jakobnielsenUsabilityEngineering1993,donalda.normanEmotionalDesignWhy2004,berndh.schmittExperientialMarketingHow2000,russellw.belkExtendedSelfDigital2013,russellw.belkPossessionsExtendedSelf1988}.}
\revise{This broader scope is also reflected in HCI-related discussions of customer experience (CX), where CX has been examined as an extension of UX and as a bridge between service science and human--computer interaction when experience involves multiple products, systems, or services rather than a single interface ~\cite{rusuCustomerEXperienceBridge2020a}. A recent systematic review further shows that CX definitions and dimensions have been examined from an HCI perspective, while also emphasizing that CX dimensions vary across domains \cite{quinonesUnderstandingCustomerExperience2023}. 
We therefore draw on CX research as a complementary source of multidimensional experience constructs, particularly for organizing experience across products, services, brands, and touchpoints~\cite{katherinen.lemonUnderstandingCustomerExperience2016,gentileHowSustainCustomer2007,markusgahlerCustomerExperienceConceptualization2023}.}


This expanded construct space is especially relevant for smartphones.
Smartphones have become the most pervasive personal technology~\cite{anttioulasvirtaHabitsMakeSmartphone2012,fabioduarteTimeSpentUsing2025}, and their ubiquity and multifunctionality make them a salient domain for studying UX. 
Accordingly, smartphone UX research has examined technical and interactional qualities, such as OS performance, application usability, interface design, and aesthetics ~\cite{sebastiana.c.perrigSmartphoneAppAesthetics2023,jiegaoExploringSmartphoneUser2024,pawelweichbrothUsabilityIssuesMobile2025}. It has also studied usage contexts and tasks, such as mobile payment, photography, entertainment, and social interaction~\cite{anttioulasvirtaHabitsMakeSmartphone2012,markdereuverDomesticationSmartphonesMobile2016,leehumphreysMobileSocialMedia2013}, as well as broader psychological and social outcomes, including satisfaction, loyalty, identity expression, dependency, and cross-cultural differences ~\cite{lucapetruzzellisMobilePhoneChoice2010,erkanbayraktarMeasuringEfficiencyCustomer2012a,liuguohaPerceivedImpactProduct2025,weifenghuResearchBrandImage2023,chengjunzhouPaperResearchMobile2024}. 
\revise{Taken together, this literature suggests that smartphones should be understood not only as tangible devices, but also as operating-system platforms, gateways to digital services, ecosystem nodes, brand carriers, and socially visible personal objects. 
A smartphone UX framework therefore needs to account for functional use, sensory interaction, and social or symbolic meanings within a common measurement structure.}

Reflecting this breadth, prior work has operationalized smartphone or mobile UX through multiple measurement traditions that differ in how they decompose experience and where they set measurement boundaries~\cite{ehsanmortazaviExploringLandscapeUX2024}. 
Mobile-phone usability measurement has been approached both through structured, checklist-based evaluation frameworks~\cite{jeongyunheoFrameworkEvaluatingUsability2009} and through validated questionnaires for comparative assessment of phone models and prototypes~\cite{youngsamryuReliabilityValidityMobile2006,youngryuDecisionModelsComparative2007}. 
Complementary strands move beyond pure usability by introducing holistic mobile UX instruments that emphasize mobility-specific facets (e.g., nuisance, mobility, access) and target software–device use scenarios~\cite{soussandjamasbiMUXDevelopmentHolistic2017}, as well as industry-oriented frameworks~\cite{jiatanFrameworkSoftwareUsability2013}. 
At the same time, mobile UX is often measured at the app level rather than at the operating system (OS) level~\cite{jeffsaurophdSUPRqmQuestionnaireMeasure2017,janvanderlindenUserExperienceUX2025}.
\revise{
These frameworks provide important foundations, but they do not directly offer a smartphone-oriented hierarchy that connects broad experience constructs with fine-grained UX metrics observable in user-generated reviews.
Table~\ref{tab:ux-framework-comparison} compares representative frameworks against four requirements of review-based smartphone UX measurement. These requirements are theoretical breadth, smartphone-domain specificity, hierarchical interpretability, and review-based operationalizability.
This comparison situates the proposed hierarchy within existing work and clarifies its task-specific role. Existing UX theories and instruments provide conceptual foundations. Our framework adapts these foundations into a domain-specific measurement structure that links theory-grounded experience constructs with fine-grained evidence expressed in smartphone reviews.
}

\begin{table*}[ht]
\caption{\revise{Comparison of representative experience frameworks with the requirements of review-based smartphone UX measurement.}}
\label{tab:ux-framework-comparison}
\centering
\small
\renewcommand{\arraystretch}{1.15}
\begin{tabularx}{\textwidth}{
>{\raggedright\arraybackslash}p{0.20\textwidth}
>{\raggedright\arraybackslash}p{0.18\textwidth}
>{\raggedright\arraybackslash}X
>{\raggedright\arraybackslash}X
}
\toprule

\revise{\textbf{Framework or instrument}}
& \revise{\textbf{Primary focus}}
& \revise{\textbf{Contribution to this study}}
& \revise{\textbf{Gap for our task}} \\

\midrule

\revise{ISO 9241-210~\cite{internationalorganizationforstandardizationErgonomicsHumansystemInteraction2019}}
& \revise{General UX definition}
& \revise{Defines UX broadly as users' perceptions and responses to a product, system, or service}
& \revise{Does not specify smartphone-specific categories, metric boundaries, or review coding units} \\

\revise{Pragmatic--hedonic UX models~\cite{marchassenzahlThingUnderstandingRelationship2003,marchassenzahlNeedsAffectInteractive2010}}
& \revise{Pragmatic, hedonic, affective, and self-related qualities}
& \revise{Grounds the distinction between instrumental and experiential qualities}
& \revise{Does not provide a smartphone-device-level hierarchy for user review analysis} \\

\revise{Mobile-phone usability and comparative assessment~\cite{youngsamryuReliabilityValidityMobile2006,youngryuDecisionModelsComparative2007,jeongyunheoFrameworkEvaluatingUsability2009}}
& \revise{Mobile-phone usability and comparative evaluation}
& \revise{Informs functional metrics related to use, operation, and interaction quality}
& \revise{Provides limited guidance for sensory, social, symbolic, ecosystem-related, and brand-mediated experience} \\

\revise{Holistic mobile UX instruments~\cite{soussandjamasbiMUXDevelopmentHolistic2017}}
& \revise{Mobile UX and mobility-specific facets}
& \revise{Extends mobile evaluation beyond narrow usability and highlights mobile use conditions}
& \revise{Often targets mobile use scenarios or software--device interaction rather than whole-device smartphone reviews} \\

\revise{App-level UX instruments and app-review studies~\cite{jeffsaurophdSUPRqmQuestionnaireMeasure2017,janvanderlindenUserExperienceUX2025}}
& \revise{App-level UX, software issues, and user evaluations}
& \revise{Supports operational thinking for analyzing user-generated reviews}
& \revise{Focuses on applications or software issues rather than smartphone-device-level UX constructs} \\

\revise{CX theories and multidimensional CX scales~\cite{katherinen.lemonUnderstandingCustomerExperience2016,gentileHowSustainCustomer2007,markusgahlerCustomerExperienceConceptualization2023}}
& \revise{Cross-touchpoint and multidimensional customer experience}
& \revise{Provides a broad construct space including cognitive, physical, sensorial, symbolic, relational, and affective responses}
& \revise{Does not directly specify smartphone-specific or review-observable metric boundaries} \\

\revise{Smartphone satisfaction and brand-related research~\cite{lucapetruzzellisMobilePhoneChoice2010,erkanbayraktarMeasuringEfficiencyCustomer2012a,weifenghuResearchBrandImage2023}}
& \revise{Satisfaction, perceived value, loyalty, and brand image}
& \revise{Informs evaluative, consumer-facing, and social-symbolic interpretation}
& \revise{Often treats experience as outcome variables rather than fine-grained, hierarchical UX metrics} \\

\bottomrule
\end{tabularx}
\end{table*}

\subsection{Measuring UX from Online Reviews and Naturalistic Discourse}

In the digital era, the abundance of online user reviews provides an unprecedented opportunity to observe and study users experience \cite{fangminchengUserExperienceEvaluation2021}. Compared with traditional surveys or interviews, user reviews offers high temporal resolution, large scale, and naturalistic context \cite{kevinlanekellerStrategicBrandManagement2020}, making it an ideal data source for capturing the dynamic layer of users' perceptions \cite{whitney-jocelynkouahoInvestigatingPerspectivesExperiences2024,mohammedaldeenEndusersKnowBest2024,lucapajolaNovelReviewHelpfulness2023,waltert.nakamuraWhatFactorsAffect2022}. 
In the smartphone domain, review-based studies have leveraged such discourse to characterize experience at multiple levels of abstraction and for diverse analytic goals. One line of work focuses on experience representation, extracting product features or aspects from reviews and organizing them into structured representations (e.g., feature structure trees or prioritized feature sets), sometimes supported by reusable aspect inventories from curated datasets \cite{xinshengxuApproachExtractProduct2017,lilishiProductFeatureExtraction2023,salemalghamdiAspectbasedSentimentAnalysis2024}. A second stream operationalizes experience evaluation and decision support by combining aspect identification with sentiment or satisfaction estimates (often with importance weighting) to profile strengths and pain points, construct evaluation indices, and derive improvement, recommendation, or competitive insights \cite{chengyangOnlineUserReview2021,huizhangProductInnovationBased2018,jihuacaoOnlineReviewsSentiment2023,yuwangProductCompetitiveAnalysis2025}. Finally, reviews and broader UGC have been used to infer brand and social meaning, including brand positioning/image via review mining and association-based scoring, as well as community-related outcomes in smartphone brand communities \cite{satanikmitraOBIMComputationalModel2020,heUnderstandingConsumersMulticompeting2023,nanfengEffectsReviewSpam2018}.

While online reviews provide rich, naturalistic evidence of the smartphone experience, many review-based analyses focus on feature or aspect summarization or other task-driven outcomes \cite{yuanchunliMiningUserReviews2017}. Although topics are useful analytic units, they do not necessarily align with a theory-grounded UX construct space, and review mining outputs therefore often provide limited support for organizing experience into stable, reusable UX metrics. Moreover, results are commonly reported as task-specific artifacts and are not primarily structured as hierarchical measurement representations that enable consistent aggregation and comparison across products and settings.


\subsection{\revise{From Opinion Mining to Construct-Grounded UX Measurement}}

\revise{Transforming naturalistic reviews into UX measurement requires computational methods, but the goal differs from generic opinion mining. For smartphone UX measurement, extracted labels must correspond to stable experience constructs, preserve the evidence behind local review expressions, and support aggregation across brands, platforms, and metric levels.}

Leveraging unstructured user reviews for reliable review-based UX measurement still faces multiple challenges, including data heterogeneity across platforms~\cite{gangkouCrossplatformMarketStructure2021,renevieirasantinSystematicReviewAspectbased2025}, semantic ambiguity in complex Chinese contexts~\cite{haiyunpengReviewSentimentAnalysis2017}, and the frequent presence of mixed emotions within single comments~\cite{kexuanniuEventawareSarcasmDetection2025,lalehdavoodiAutomatingCustomerFeedback2026}. A large body of work aggregates raw review text into stable conclusions through opinion mining and aspect-based sentiment analysis (ABSA)~\cite{tunthurathetAspectbasedSentimentAnalysis2010,bingliuSentimentAnalysisOpinion2012,mariapontikiSemEval2016Task52016}.

Typical pipelines formulate outputs such as extracted attributes, associated opinion expressions, and sentiment polarity or graded satisfaction~\cite{wenxuanzhangSurveyAspectbasedSentiment2023}, enabling downstream aggregation across many reviews. 
In practice, many prior studies thus summarize attitudes using simplified indicators~\cite{rahularalikatteFaultYourStars2018,fenghuMappingHotelBrand2020} or derive brand-related associations via text-mining techniques~\cite{odednetzerMineYourOwn2012,argahanantoApplicationTextMining2016,nimashaarambepolaFactorsInfluencingMobile2024}.
In recent years, large language models (LLMs) have demonstrated notable advantages across a range of information extraction~\cite{derongxuLargeLanguageModels2024,johndagdelenStructuredInformationExtraction2024} and sentiment analysis tasks~\cite{wenxuanzhangSentimentAnalysisEra2024}, including extracting structured fields from unstructured text and handling implicit or context-dependent expressions~\cite{chaoxupangGuidelineLearningIncontext2023}. 
In fine-grained aspect-based sentiment analysis~\cite{tunthurathetAspectbasedSentimentAnalysis2010,bingliuSentimentAnalysisOpinion2012,mariapontikiSemEval2016Task52016}, LLMs are now capable of identifying complex expressions that convey mixed sentiments~\cite{divyamsobtiDomainspecificAspectExtraction2025}. LLMs have been applied to define and extract attributes from e-commerce reviews~\cite{andrerusliLeveragingLLMsAttribute2024,anselblumeGenerativeModelsProduct2023}, perform multi-dimensional fine-grained sentiment scoring \cite{bhaveshkukrejaSentimentMultifacetedReview2024}, analyze competitor user reviews~\cite{maramassiLLMcureLLMbasedCompetitor2025}, and support generative multi-modal attribute extraction~\cite{anantkhandelwalLargeScaleGenerative2023}.
Meanwhile, LLMs have advanced explorations in HCI, psychology, and sociology, including context-aware models based on the theory of constructed emotion~\cite{nilskluwerContextCategoriesImplementing2025}, and the system supporting UX metric construction~\cite{qingxiaozhengEvAlignUXAdvancingUX2025}.
These capabilities make LLM-based methods attractive for review understanding, where experience descriptions may be indirect, multi-faceted, or rhetorically compressed~\cite{zengzhiwangChatGPTGoodSentiment2023}. At the same time, applying LLMs in review analytics foregrounds the need to control output formats and ensure consistent interpretations so that extracted structures can be aggregated reliably~\cite{leihuangSurveyHallucinationLarge2025,saibogengJSONSchemaBenchRigorousBenchmark2025}.

Despite these advances, existing methods often prioritize extraction or classification accuracy within fixed datasets, while offering limited support for stable evidence schemas and consistent label interpretation. Sentiment inference in reviews is also sensitive to context qualification and mixed emotions, which complicates calibrated comparison across brands or platforms.

\revise{Taken together, prior work provides important foundations but does not yet offer an integrated basis for smartphone-oriented, review-based UX measurement. UX and CX research shows that experience extends beyond usability and satisfaction, but existing frameworks are often too general, too outcome-oriented, or insufficiently operationalized for smartphone-specific review analysis. Smartphone review-mining studies can extract aspects, topics, and sentiments at scale, but their outputs are often organized around surface product features or task-specific artifacts rather than stable UX constructs. LLM-based methods further improve the ability to interpret implicit and multi-faceted review expressions, but they also require explicit schemas to ensure consistent labeling and meaningful aggregation. These gaps motivate our approach: a hierarchical smartphone UX measurement framework that connects theory-grounded constructs with review-observable evidence and supports interpretable comparison across brands and platforms.
}


\section{Developing the Hierarchical Smartphone UX Measurement Framework}
\label{3}
This section describes how we developed the hierarchical smartphone UX measurement framework used in the subsequent computational analysis. We aimed to construct a smartphone-oriented measurement specification that could support both conceptual interpretation and large-scale review-based analysis. The framework therefore needed to cover the main experiential qualities through which users evaluate smartphones and provide metric boundaries that could be identified in naturally occurring online comments. We developed the framework iteratively by reviewing theoretical models and prior studies from HCI, psychology, and consumer research, adapting the initial construct space to smartphone use, and refining the hierarchy through expert review and examination of publicly accessible cross-platform user comments. 
The study protocol was reviewed and approved by the institutional ethics committee.

\subsection{Initial Metric Hierarchy Development}
\label{3.1}

\revise{
To develop a smartphone-oriented UX measurement framework that was both theory-grounded and operable for user-generated reviews, we constructed the initial metric hierarchy through a deductive--inductive process. The hierarchical structure was used because smartphone review evidence varies in granularity. Broad constructs are needed to organize experience at a stable conceptual level, while lower-level metrics are needed to capture concrete issues expressed in reviews. We first defined an upper-level construct space from UX and CX literature~\cite{markusgahlerCustomerExperienceConceptualization2023,kevinlanekellerStrategicBrandManagement2020,fredd.davisPerceivedUsefulnessPerceived1989}. We then adapted these constructs to smartphone use and refined lower-level metrics through recurrent expressions observed in user-generated smartphone reviews. This process connected high-level experience constructs with review-observable metrics before the panel-based review described in Section~\ref{3.2}.
}


\revise{
Building on the discussion of multidimensional UX and complementary CX perspectives in Section~\ref{2.1}, we used the six-dimensional, 18-item Customer Experience (CX) scale proposed by Gahler et al.~\cite{markusgahlerCustomerExperienceConceptualization2023} as an initial construct reference. The scale was useful because it offers a recent measurement-oriented structure for organizing multidimensional experience, including affective, cognitive, physical, relational, sensorial, and symbolic dimensions. Because our goal was to construct a taxonomy for review-based smartphone UX measurement, we used these dimensions as conceptual starting points and adapted them to smartphone-domain evidence rather than applying the original item structure directly.
}

\revise{
The adaptation was necessary because smartphone reviews often describe experiences that span device properties, software and service use, ecosystem relations, brand meanings, and socially visible ownership. We therefore reorganized the CX construct space according to the experiential roles expressed in smartphone use and review evidence.}
\revise{We reorganized the construct space into three Level-1 categories. Functional Experience covers goal support, performance, effort, reliability, durability, and task completion. Sensory Experience covers visual, auditory, tactile, material, and thermal perception. Social Experience covers identity expression, brand association, ecosystem belonging, interpersonal connection, community participation, and social acceptance. In this framework, Social Experience captures socio-symbolic and relational meanings around smartphone ownership and use, beyond communication functions alone.
}

\revise{
We treated affective response differently from the three Level-1 attribute categories. The hierarchy identifies what users evaluate, while affective expressions in reviews usually indicate how users evaluate those experiences. Because sentiment polarity is modeled separately in the computational pipeline, including affective response as a parallel experience attribute could mix experience objects with evaluative valence.
}

\revise{
After defining the three Level-1 categories, we specified lower-level metrics by combining literature-based construct anchoring with smartphone-domain adaptation. Level-2 metrics represent relatively stable construct families grounded in UX, technology acceptance, product evaluation, branding, and mobile experience research~\cite{fredd.davisPerceivedUsefulnessPerceived1989,philipkotlerMarketingManagement2016,jonahbergerWordMouthInterpersonal2014}. Level-3 metrics translate these construct families into smartphone-domain subconstructs~\cite{alibabaclouduxteamEaseUseMetric2020,kevinlanekellerConceptualizingMeasuringManaging1993,kevinlanekellerStrategicBrandManagement2020}. Level-4 metrics capture concrete expressions that appear recurrently in user comments and provide more specific diagnostic information.
}

\revise{
Recurrent themes were first identified through clustering of user comments and then assigned conceptual labels with reference to established constructs and smartphone-domain interpretation. During this inductive refinement process, recurring UX-related expressions that could not be accommodated by the existing hierarchy were examined as candidates for adding or revising metrics. Expressions unrelated to product experience, such as promotion, logistics, resale, purchase inquiries, or general brand discussion, were excluded from the UX measurement scope. In this process, Level-3 metrics served as conceptual anchors with relatively stable semantic boundaries, while Level-4 metrics captured more concrete, data-grounded experiential expressions.
}

\revise{
The resulting initial hierarchy was a domain-adapted measurement specification. It was built from existing UX and CX constructs, but adapted to smartphone-domain evidence and review-based operationalization. The hierarchy links broad experience categories to mid-level constructs and fine-grained review-observable metrics. It also provides the construct space for the subsequent computational pipeline, in which review expressions are extracted, mapped to the hierarchy, assigned valence scores, and aggregated for cross-brand and cross-platform analysis.
}

\begin{table}[ht]
\caption{The hierarchical measurement framework for smartphone user experience (L1-L3).}
\label{tab:framework}
    \centering
    
\small  
\begin{tabular}{|l|l|l|}
\hline
\rowcolor[HTML]{C0C0C0} 
\textbf{Level-1 metrics}                         & \textbf{Level-2 metrics}                      & \textbf{Level-3 metrics}            \\ \hline
                                             &                                            &  Time Saving                       \\ \cline{3-3} 
                                             &                                            &
                                             Labor Saving                   \\ \cline{3-3} 
                                             & \multirow{-3}{*}{Usefulness}           & Increase Usage Benefits         \\ \cline{2-3} 
                                             &                                            & Learnability                    \\ \cline{3-3} 
                                             &                                            & Clarity                         \\ \cline{3-3} 
                                             & \multirow{-3}{*}{Ease of Use}         & Operability                     \\ \cline{2-3} 
                                             &                                            & Stability                        \\ \cline{3-3} 
                                             &                                            & Security                        \\ \cline{3-3} 
                                             & \multirow{-3}{*}{Reliability}          & Responsiveness                  \\ \cline{2-3} 
                                             &                                            & Battery Life                    \\ \cline{3-3} 
                                             &                                            & Memory Management               \\ \cline{3-3} 
\multirow{-12}{*}{Functional Experience (FE)} & \multirow{-3}{*}{Durability}           & Performance Retention           \\ \hline
                                             &                                            & Appearance and Design           \\ \cline{3-3} 
                                             &                                            & Display Quality                 \\ \cline{3-3} 
                                             &                                            & Image Aesthetics                \\ \cline{3-3} 
                                             & \multirow{-4}{*}{Visual Appeal}        & Animation Effects               \\ \cline{2-3} 
                                             &                                            & Volume                          \\ \cline{3-3} 
                                             &                                            & Sound Quality                   \\ \cline{3-3} 
                                             &                                            & Call Audio Quality              \\ \cline{3-3} 
                                             &                                            & Media Audio Quality             \\ \cline{3-3} 
                                             & \multirow{-5}{*}{Auditory Effects}     & System Audio Quality            \\ \cline{2-3} 
                                             &                                            & Haptic Feedback Effect          \\ \cline{3-3} 
                                             &                                            & Thermal Perception              \\ \cline{3-3} 
                                             &                                            & Touch Response                  \\ \cline{3-3} 
\multirow{-13}{*}{Sensory Experience (SEN)}   & \multirow{-4}{*}{Tactile Effects}      & Surface Texture                 \\ \hline
                                             &                                            & Personal Style Expression       \\ \cline{3-3} 
                                             &                                            & Brand–Self Connectedness        \\ \cline{3-3} 
                                             & \multirow{-3}{*}{Self-Expression}      & Self-Improvement                \\ \cline{2-3} 
                                             &                                            & Interpersonal Connectedness     \\ \cline{3-3} 
                                             &                                            & Community Participation         \\ \cline{3-3} 
\multirow{-6}{*}{Social Experience (SOC)}      & \multirow{-3}{*}{Social Connectedness} & Social Acceptance and Belonging \\ \hline
\end{tabular}

\end{table}

\subsection{
\revise{Framework Refinement through Panel-Based Review
}}
\label{3.2}

\revise{To refine the initial metric hierarchy, we conducted a Delphi-style panel review. The review examined the conceptual clarity, hierarchical coherence, construct boundaries, and operational applicability of the hierarchy for review-based smartphone UX analysis.}


\subsubsection{Participants}

\revise{The review panel involved 13 participants in two stages. In the first stage, ten academic panelists participated in eight Delphi-style rounds, following the Delphi tradition of iterative expert judgment and feedback~\cite{linstoneDelphiMethodTechniques1975}. This stage focused on refining the conceptual structure, metric definitions, hierarchical relationships, and construct boundaries of the framework. In the second stage, three smartphone industry practitioners participated in a final-round review to assess the practical relevance, product-facing interpretability, and actionability of the revised framework.}

\revise{The academic panel was assembled through purposive sampling. The selection prioritized task-relevant expertise and complementary perspectives across the theoretical, methodological, and domain-specific requirements of the framework. Panelists had research or project experience related to experience evaluation, consumer behavior, technology use, construct interpretation, qualitative analysis, taxonomy refinement, or smartphone product experience. This composition allowed the review to combine conceptual and methodological oversight with detailed feedback on metric definitions, construct boundaries, and the applicability of the hierarchy to user-generated smartphone reviews.}

\revise{The selection strategy also helped reduce the risk that the refinement process would be shaped by a single disciplinary perspective. The academic panel reviewed the framework from complementary perspectives, including the adequacy of experience constructs, the clarity of metric definitions and boundaries, and the applicability of the hierarchy to smartphone review analysis.}

\revise{The three industry practitioners were recruited from the smartphone sector. Their review focused on whether the revised hierarchy, metric names, and definitions were interpretable and actionable in product-facing smartphone UX analysis. Their feedback was used to refine terminology, examples, and practical interpretation of the metrics. Additional anonymized information about the panelists' roles, relevant background, and contributions to the review is provided in Appendix~\ref{Experts Information}.}

\subsubsection{Procedure}

The study was conducted after obtaining institutional ethics approval. 
Before participation, all participants were given information about the review procedure, data handling, and their right to withdraw, and all provided informed consent.
\revise{
 The panelists reviewed the initial hierarchy developed in Section~\ref{3.1}. The review materials included metric names, definitions, hierarchical relationships, and example review evidence. The procedure was designed to refine the hierarchy and examine whether it was conceptually clear, internally coherent, and applicable to user-generated smartphone reviews. 
}

\revise{
The first stage consisted of eight Delphi-style rounds with the academic panel. In each round, panelists reviewed the current version of the hierarchy and provided written comments on metric coverage, hierarchical assignments, construct boundaries, and definition clarity. The early rounds examined the overall coverage of the Level-1 to Level-3 structure and whether important smartphone UX dimensions were missing. The middle rounds focused on overlaps, ambiguities, and hierarchical inconsistencies among metrics. The later rounds refined metric names, definitions, examples, and coding-oriented descriptions so that the framework could be operationalized on user-generated reviews. In the final academic round, panelists reviewed the revised hierarchy and provided confirmation or remaining comments on unresolved issues.
}

\revise{
After each round, the research team summarized the feedback and grouped comments into four categories: coverage-related comments, hierarchy-related comments, construct-boundary comments, and operational-clarity comments. Revisions were made by adding, merging, splitting, renaming, or redefining metrics. When panelists raised conflicting suggestions, the research team compared the comments with the theoretical foundations described in Section~\ref{3.1} and the task-specific requirements summarized in Section~\ref{2.1}. We prioritized revisions that improved conceptual clarity, reduced category overlap, preserved hierarchical interpretability, and maintained applicability to smartphone review analysis. The revised hierarchy, together with a summary of major changes, was circulated to the panelists in the following round for further review.}
\revise{
As part of this iterative process, we also examined whether recurring UX-related expressions in the collected reviews could be represented by the current hierarchy. Expressions that could not be accommodated were discussed as candidates for adding or revising metrics. Non-UX content, such as promotion, logistics, resale, purchase inquiries, and general brand discussion, was treated as outside the measurement scope. Major revisions were re-examined by the panel in subsequent rounds before being incorporated into the finalized hierarchy.
}

\revise{
We considered the adequacy of the panel size in relation to the intended scope of the review. Consistent with content-validity and Delphi-method literature, we evaluated panel adequacy based on reviewer relevance, purposive selection, structured iteration, and convergence of feedback, instead of statistical representativeness~\cite{haynesContentValidityPsychological1995,chituokoliDelphiMethodResearch2004,hsuDelphiTechniqueMaking}. The review combined a purposively assembled panel, eight rounds of iterative feedback, and repeated re-examination of revised metrics. By the final rounds, the review had reached practical convergence. Panel comments no longer suggested adding recurring UX-related metrics or substantially restructuring the hierarchy. No major unresolved disagreement remained regarding metric placement or construct boundaries, and remaining comments mainly concerned wording, examples, or local boundary clarification. We treated this as corpus-grounded adequacy within the studied review context. It was not treated as evidence of universal completeness.
}

\revise{
In the second stage, three smartphone industry practitioners reviewed the revised hierarchy. This final-round review focused on the practical relevance, product-facing interpretability, and actionability of the metric names and definitions in smartphone UX analysis. Their feedback was used to refine terminology and examples. It was not used to re-establish the conceptual structure of the hierarchy.
}
\revise{
Accordingly, the panel review provides content-oriented and face-validity evidence for the hierarchy. It supports the framework's content coverage, conceptual clarity, hierarchical coherence, construct-boundary clarity, and operational applicability to review-based smartphone UX analysis. Computational performance and contextual transferability are examined separately in the subsequent sections.
}

\section{\revise{User Review Corpus and Computational Operationalization of Smartphone UX}}

Building on the hierarchical smartphone UX measurement framework introduced in Section~\ref{3}, this chapter describes how we operationalize the conceptual metrics into a computational pipeline for large-scale analysis of cross-platform user reviews. Our goal is to translate naturalistic review text into localized, metric-labeled evidence units with comparable satisfaction scores, enabling structured aggregation and comparison across brands and platforms. We first introduce the end-to-end pipeline (Section~\ref{4.1}), and then detail the data sources (Section~\ref{4.2}), preprocessing procedures (Section~\ref{4.3}), task formulations (Section~\ref{4.4}), and implementation and evaluation (Section~\ref{4.5}).

\subsection{Pipeline Overview}
\label{4.1}

\begin{figure}[ht]
    \centering
    \includegraphics[width=0.75\textwidth]{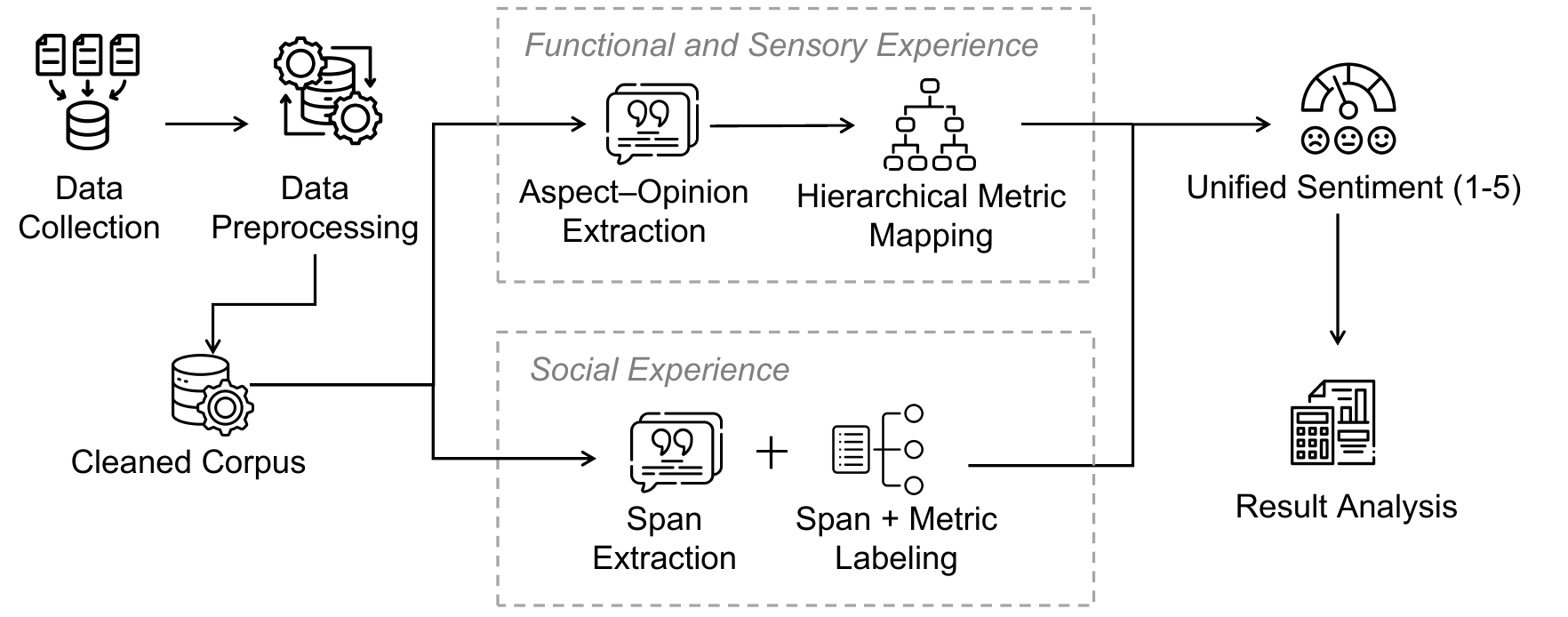}
    \caption{Pipeline overview of our computational operationalization of the hierarchical smartphone UX metrics.}
    \label{fig:1}
\end{figure}

\revise{To operationalize the hierarchical framework at scale, we developed an end-to-end pipeline that transforms user reviews into structured UX evidence through evidence extraction, hierarchical metric mapping, and unified satisfaction scoring. 
As shown in Figure~\ref{fig:1}, the collected reviews are first preprocessed to construct a cleaned corpus. 
Functional and Sensory Experiences are represented as aspect--opinion pairs and mapped to the metric hierarchy through coarse-to-fine classification.
For Social Experience, contextual spans are identified and jointly mapped to the corresponding metrics.
A unified five-level satisfaction classifier then assigns a satisfaction score to each metric-labeled evidence unit. 
The resulting metric--satisfaction records form the basis for the cross-brand and cross-platform analyses presented in Section~\ref{5}.}

\subsection{\revise{Review Corpus and Sampling Context}}
\label{4.2}
\revise{To operationalize the proposed framework on naturally occurring smartphone discourse,}
we constructed a cross-platform corpus of smartphone-related user reviews from major Chinese social media platforms, including Weibo, RedNote, and Bilibili. The dataset was assembled through our industry collaboration and consisted of publicly accessible user comments about flagship-series smartphones.
\revise{These platforms capture different review genres. Weibo mainly provides short opinion-oriented posts, RedNote provides narrative lifestyle reviews, and Bilibili provides comments around video reviews and technical demonstrations.}
Combining these platforms allowed us to capture a broader range of user groups, review genres, and linguistic styles, thereby \revise{broadened the range of observable review genres and reduced reliance on any single platform.}


For compliance with our industry partner's requirements, we anonymize the four brands as Brand A-D throughout the paper. The mapping is fixed across all analyses and figures.
These brands were selected from flagship-series smartphones in the Chinese market to cover different operating systems, market segments, and brand positions, while ensuring sufficient volumes of user discussion across platforms. 
\revise{The collection covered one full product launch cycle and yielded approximately 100,000 comments.}
The analysis was restricted to publicly accessible comments and involved no direct interaction with posters. \revise{We removed direct identifiers such as usernames, profile links, and URLs, and paraphrased reported examples to reduce traceability.}

\subsection{Data Preprocessing}
\label{4.3}

To ensure the validity of the subsequent opinion extraction and sentiment analysis, we \revise{applied text-based preprocessing to the raw cross-platform corpus.}
\revise{We removed promotional, templated, and non-opinionated comments using text-based heuristic rules.}
\revise{After filtering, 20,837 reviews were retained for the corpus-level analysis, with variation across brands and platforms reflecting differences in initial collection volume and filtering outcomes (Table~\ref{tab:data}). Each retained review preserved a unique original-review identifier, allowing all UX evidence units derived from the same review to remain linked during subsequent model evaluation.}

\begin{table}[htbp]
\caption{Number of reviews retained for each brand.}
\label{tab:data}
\small
\begin{tabular*}{0.6\textwidth}{@{\extracolsep\fill}lccccc}
\toprule%
Platform & Brand A & Brand B & Brand C & Brand D \\
\midrule
Weibo    & 1083 & 1530  & 2030   & 1510  \\
RedNote  & 1636 & 2057 & 1751   & 2232  \\
Bilibili & 2062 & 1417 & 1749   & 1780  \\
Total    & 4781 & 5004 & 5530   & 5522   \\
\bottomrule
\end{tabular*}

\end{table}

\subsection{\revise{Computational Operationalization of UX Evidence}}
\label{4.4}

\subsubsection{\revise{Evidence Identification and Metric Mapping}}

\revise{The three experience domains differ in how they appear in user reviews. FE and SEN are often expressed through concrete system attributes, such as battery, screen, or sound, together with evaluative opinions. SOC is more often embedded in contextual statements about identity expression, social relationships, and belonging. We therefore used domain-specific evidence representations while mapping all outputs to the same hierarchical UX framework.}

For FE and SEN, we employed a large language model (GPT-4o-mini) to extract (aspect, opinion) pairs from raw review texts, which serve as the basic evidence units for subsequent modeling. 
Aspect–opinion extraction is applied only when sufficient linguistic evidence is present. Reviews lacking explicit functional or sensory references are excluded from this pipeline.
The prompt for the extraction task is detailed in Appendix~\ref{prompt:extrcation}.

\revise{The extracted FE and SEN evidence units were mapped using fine-tuned ERNIE-3.0-Xbase-zh classifiers \cite{baidu-ernie-teamERNIE45Technical2025, paddlenlpcontributorsPaddleNLPEasytouseHigh2021}. 
Rather than applying a single flat classifier across all lower-level metrics, we used a coarse-to-fine hierarchical procedure. 
Each unit was first assigned to an L2 metric and was then classified within the corresponding L3/L4 branch. 
This design addresses uneven label distributions and semantic overlap by restricting each lower-level prediction to the relevant parent branch.}

\revise{SOC evidence is less often expressed through explicit system attributes and is more dependent on surrounding context. We therefore used GPT-4o-mini to identify semantically coherent spans related to SOC. For each relevant review, the model extracted one or more spans and assigned each span to a Level-3 SOC metric. This retained the contextual information needed to interpret implicit SOC expressions. The prompt is provided in Appendix~\ref{prompt:span}.}


\subsubsection{Unified Satisfaction Sentiment Classification}
To represent the evaluative intensity associated with each UX metric, we assigned every identified evidence unit a score on a common five-level scale, ranging from strong dissatisfaction (1) to strong satisfaction (5). A score of 3 indicates a neutral or mixed evaluation, while 2 and 4 indicate mild dissatisfaction and mild satisfaction, respectively. Detailed annotation guidelines and examples are provided in Appendix~\ref{Satisfaction Guidelines}. 
\revise{A fine-tuned ERNIE-3.0-Xbase-zh model applied this scoring scheme to FE and SEN aspect--opinion pairs and SOC spans. The score represents the evaluation of a specific UX evidence unit rather than the reviewer's overall attitude toward a device. This evidence-level scoring allows a single review to express different satisfaction levels for different aspects of experience. The final analytical record contains the source review, textual evidence, hierarchical metric label, and satisfaction score.}

\subsection{Implementation and Model Evaluation}
\label{4.5}
\revise{We evaluated the computational operationalization through component-level model evaluations and an end-to-end human validation. The component evaluations examine evidence identification, hierarchical metric mapping, and satisfaction scoring, while the end-to-end validation examines complete pipeline outputs from original reviews.}


\subsubsection{Evaluation of Extraction Consistency}

To assess the reliability of the LLM-based extraction, we conducted an experiment to verify the consistency between GPT-4o-mini outputs and expert annotations. Two domain experts independently annotated a random sample of 200 reviews following the task definitions in Section~\ref{4.4}. We measured task-level matching accuracy between the GPT-4o-mini predictions and the \revise{human reference annotations.}
\revise{The two annotators first worked independently, and disagreements were resolved through discussion to establish consensus reference annotations.}
For FE and SEN aspect–opinion extraction, we computed the semantic similarity between the predicted and annotated (aspect, opinion) units, and counted a unit as matched if similarity exceeded 0.5. For social span extraction, we used span-level overlap measured by IoU, and considered a review matched if more than half of the spans in that review achieved IoU > 0.7 between the GPT-4o-mini and the consensus annotation. For social metric classification, we reported macro classification accuracy over the six social categories.
\begin{table}[ht]
\caption{Consistency between GPT-4o-mini and human experts.}
\label{tab:consistency}
 \small
\begin{tabular*}{0.5\textwidth}{@{\extracolsep\fill}lc}
\toprule%
Task                              & Matching Rate \\ 
\midrule
FE and SEN Extraction & 0.7752        \\
SOC Span Extraction            & 0.7600        \\
SOC Classification             & 0.8246        \\ 
\bottomrule
\end{tabular*}

\end{table}

GPT-4o-mini achieved strong consistency with expert consensus, as shown in Table \ref{tab:consistency}. 
The slightly lower matching rates for the FE and SEN extraction task and the SOC span extraction task are expected, as aspect boundary selection and span segmentation in user reviews can be ambiguous and context-dependent, leading to reasonable annotator variability \cite{anselblumeGenerativeModelsProduct2023,chaoxupangGuidelineLearningIncontext2023}.
\revise{These results show that the LLM-based procedure achieved reasonable task-level correspondence with the human reference annotations under the predefined matching criteria.}

\subsubsection{Evaluation of Hierarchical Metric Mapping}
\label{sec:mapping-evaluation}

\revise{
We evaluated the supervised hierarchical mapping of Functional and Sensory evidence units, including the L2 router and the branch-specific L3/L4 classifiers described in Section~\ref{4.4}. Each classifier was trained and evaluated on its corresponding manually annotated dataset.
}
\revise{
For each classifier, we performed five-fold cross-validation grouped by original review, ensuring that all evidence units derived from the same review remained in the same fold. Exact duplicate reviews and evidence units were removed before evaluation. The L2 router was evaluated independently, whereas the branch-specific classifiers were evaluated using the reference L2 parent labels.
}

\revise{
For each fold, we calculated micro-F1 and macro-F1 and then averaged the scores across the five folds. Micro-F1 summarizes overall instance-level performance, whereas macro-F1 assigns equal weight to each class and is more sensitive to uneven fine-grained class distributions. As shown in Table~\ref{tab:classification}, the L2 router achieved a micro-F1 of 0.8757 and a macro-F1 of 0.8817. 
The modest differences between the two measures indicate some variation across fine-grained classes, particularly in branches with less balanced class distributions.
}

\revise{
These results support the use of L2 as the primary level for corpus-level comparison in Section~\ref{5}, while the branch-specific results provide evidence for fine-grained drill-down once the relevant parent metric is known. Appendix~\ref{Detailed Performance} reports detailed class distributions, per-class performance, and a concise analysis of the misclassified held-out evidence units. Across the seven branch-specific classifiers, 286 evidence units were misclassified, corresponding to an overall error rate of 6.03\%. These remaining errors mainly involved compact, implicit, or context-dependent review expressions.
}


\begin{table}[ht]
\caption{Performance of the L2 router and branch-specific child-metric classifiers under review-level grouped five-fold cross-validation.}
\label{tab:classification}
 \small
 \centering
\begin{tabular*}{0.8\textwidth}{@{\extracolsep\fill}llllll}
\toprule
\revise{Classifier Component}     & \revise{Target Level} & \revise{Classes} & \revise{Evidence Units} & \revise{Micro F1} &  \revise{Macro F1} \\ 
\midrule

L2 Router                & \revise{L2  }         & \revise{7       }& \revise{ 2786    }& \revise{ 0.8757     }& \revise{ 0.8817  }    \\
Usefulness               & \revise{L3/L4}     & \revise{ 15      }& \revise{ 1443     }& \revise{ 0.9589     }& \revise{ 0.8650  }   \\
Ease of Use              & \revise{L3/L4 }       & \revise{ 11      }& \revise{ 358      }& \revise{ 0.9728     }& \revise{ 0.9524 }    \\
Reliability              & \revise{L3/L4 }       & \revise{ 14      }& \revise{ 819      }& \revise{ 0.9868     }& \revise{ 0.9688 }    \\
Durability               & \revise{L3/L4 }       & \revise{ 14      }& \revise{ 431       }& \revise{ 0.9965   }& \revise{ 0.9936  }    \\
Visual Appeal            & \revise{L3/L4 }       & \revise{ 22      }& \revise{ 1025      }& \revise{ 0.9811    }& \revise{ 0.9582 }     \\
Auditory Effects         & \revise{L3/L4 }       & \revise{ 12      }& \revise{ 322        }& \revise{ 0.9698   }& \revise{ 0.9423  }   \\
Tactile Effects          & \revise{L3/L4 }       & \revise{ 14      }& \revise{ 346       }& \revise{ 0.9763   }& \revise{ 0.9420  }    \\

\bottomrule
\end{tabular*}
\end{table}


\subsubsection{Evaluation of Unified Satisfaction Sentiment Classification}

\revise{
We evaluated the unified satisfaction-scoring model on 3,082 manually annotated evidence units covering Functional, Sensory, and Social Experiences. Each unit was labeled on the five-level satisfaction scale described in Section~\ref{4.4}. 
We followed the same exact-deduplication and review-level grouped five-fold cross-validation protocol described in Section~\ref{sec:mapping-evaluation}, ensuring that all evidence units derived from the same original review remained in the same fold. For each fold, we calculated micro-F1 and macro-F1 and averaged the scores across the five folds. 
The classifier achieved a micro-F1 of 0.8310 and a macro-F1 of 0.8323, indicating comparable overall and class-balanced performance. 
}


\subsubsection{\revise{End-to-End Human Validation}}
\label{sec:end-to-end-validation}

\revise{In addition to evaluating individual components, we conducted an end-to-end human validation on a stratified sample of original reviews. This validation examines whether the complete pipeline can transform raw user reviews into structured UX evidence through three sequential stages.}

\revise{We sampled 233 original reviews from the cleaned review corpus and applied the complete pipeline to each sampled review. The sample was stratified by smartphone brand and social media platform, while also covering diverse predicted metric categories, satisfaction levels, and both single-unit and multi-unit reviews. Reviews for which the pipeline produced no output were also included, so that the annotators could check whether the review contained no extractable UX evidence or whether relevant evidence had been missed.}
\revise{The validation annotations were conducted by two researchers following predefined guidelines. The primary annotator compared the pipeline outputs with the corresponding original reviews and judged the correctness of each stage. A second researcher reviewed a subset of the annotated cases to check consistency and resolve ambiguous judgments.}

\revise{For extraction, the annotator judged whether each extracted opinion unit/span was valid and supported by the review text, and also recorded human-identifiable UX evidence units/spans missed by the pipeline. For valid extracted units/spans, metric mapping was judged by whether the predicted metric matched the meaning of the evidence, and satisfaction scoring was judged by whether the predicted score reflected the expressed attitude. Metric mapping and satisfaction scoring were evaluated only on valid extracted units/spans, because invalid or missed units/spans do not have directly evaluable downstream labels.}

\revise{We report extraction precision, recall, and F1-score to capture both invalid outputs and missed evidence. Precision was calculated over all pipeline-extracted units/spans, whereas recall was calculated over all human-identified UX evidence units/spans. We also report a strict unit/span-level end-to-end success rate. Its denominator includes all pipeline-extracted units/spans and all missed human-identified UX evidence units/spans; a case was counted as successful only when extraction, metric mapping, and satisfaction scoring were all correct.}

\revise{The results are reported in Table~\ref{tab:e2e_results}. Extraction achieved a precision of 0.9247, a recall of 0.9069, and an F1-score of 0.9157. Conditional on the 516 valid extracted units/spans, metric mapping and satisfaction scoring reached accuracies of 0.9205 and 0.9031, respectively. For strict end-to-end success, the evaluation considered both pipeline-extracted units/spans and missed human-identified units/spans. At the full-pipeline level, 441 of the 611 unit/span cases were correctly processed across all three stages, yielding an end-to-end success rate of 0.7218. These results provide direct human evidence for the quality of the structured UX records used in aggregate analysis, while also showing how errors may accumulate across sequential processing stages.}


\begin{table}[ht]
\caption{\revise{End-to-end human validation results.}}
\label{tab:e2e_results}
\centering
\small
\setlength{\tabcolsep}{5pt}
\begin{tabular*}{0.78\linewidth}{@{\extracolsep\fill}llcc}
\toprule
\revise{Stage} & \revise{Metric} & \revise{$n/N$} & \revise{Score} \\
\midrule
\revise{Extraction} & \revise{Precision} & \revise{516/558} & \revise{0.9247} \\
\revise{Extraction} & \revise{Recall} & \revise{516/569} & \revise{0.9069} \\
\revise{Extraction} & \revise{F1-score} & \revise{--} & \revise{0.9157} \\
\revise{Metric mapping} & \revise{Accuracy} & \revise{475/516} & \revise{0.9205} \\
\revise{Satisfaction scoring} & \revise{Accuracy} & \revise{466/516} & \revise{0.9031} \\
\revise{Full pipeline} & \revise{End-to-end success} & \revise{441/611} & \revise{0.7218} \\
\bottomrule
\end{tabular*}
\end{table}

\section{Characterization of Smartphone User Experience Metrics Across Brands and Platforms}
\label{5}
In this section, we examine how smartphone UX is expressed and evaluated in naturalistic online reviews across brands and platforms. 
\revise{This large-scale analysis demonstrates how the refined hierarchy can be operationalized to produce interpretable brand-, platform-, and metric-level characterizations of expressed smartphone UX.}
At a coarse level, functional and sensory metrics dominate the volume of extracted evidence, whereas social metrics tend to receive more positive evaluations.
We summarize the distribution of experience evidence in Level 2 metrics to capture salience, and quantify metric-level sentiment scores on a 1–5 scale to capture valence. 
We characterize brand-level salience and sentiment patterns using platform-aggregated results, and further stratify analyses by platform to examine how platform context is associated with shifts in what users discuss and how they evaluate these metrics. We focus on Level 2 metrics in the main results because they provide stable, theory-grounded construct boundaries for cross-brand and cross-platform comparison, while finer-grained Level-3 and Level-4 categories are more sparse and are best used for drill-down interpretation and illustrative examples, not as primary comparison units.

\subsection{Salience of L2 Experience Metrics}
\begin{figure}[ht]
    \centering
    \includegraphics[width=0.75\textwidth]{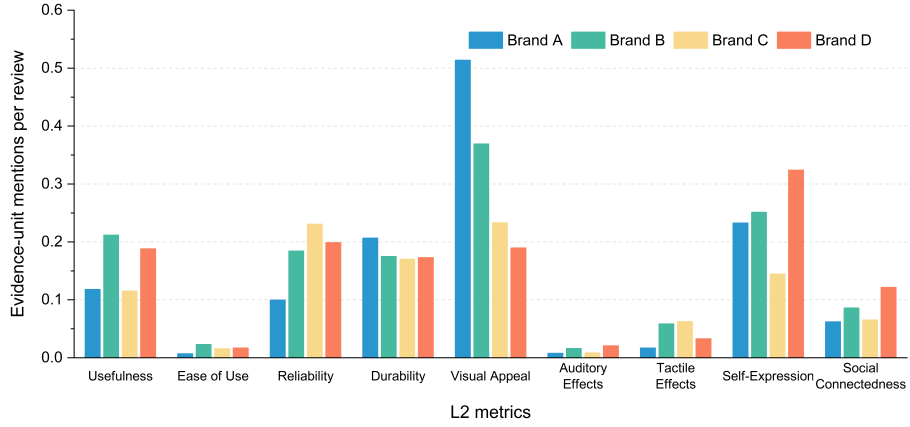}
    \caption{Brand level salience of L2 experience metrics in cross platform reviews.
}
    \label{fig:5}
\end{figure}

Figure \ref{fig:5} reports the salience of Level-2 experience metrics across brands, measured as evidence-unit mentions per review (i.e., the number of extracted evidence units mapped to each metric divided by the total number of reviews for each brand). 
Across all brands, Visual Appeal, Self-Expression and functional metrics (Reliability and Durability) show the highest normalized mention rates, indicating their high salience in real-world smartphone discussions. In contrast, Ease of Use and Auditory Effects are consistently among the least frequently mentioned metrics across all brands, suggesting that these aspects are less frequently mentioned in spontaneous user expressions.
The relative prominence of metrics also varies by brand. For example, Visual Appeal is most salient for Brand A, whereas Reliability is relatively more salient for Brand C, and Self-Expression is particularly salient for Brand D.

\subsection{Valence of L2 Experience Metrics}

\begin{figure}[ht]
    \centering
    \includegraphics[width=0.8\textwidth]{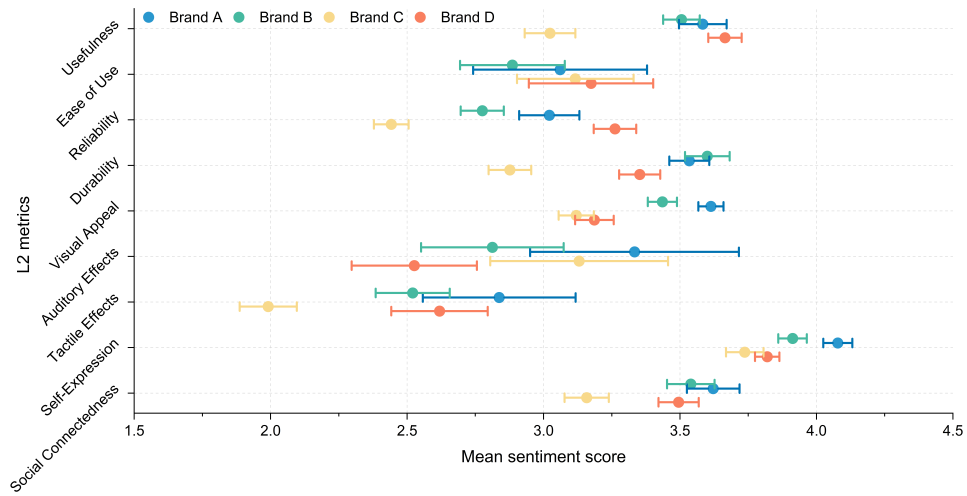}
    \caption{Brand level valence of L2 experience metrics in cross platform reviews.}
    \label{fig:6}
\end{figure}

Figure \ref{fig:6} compares the valence of L2 experience metrics across brands, operationalized as the mean evidence-unit sentiment score on a 1–5 scale for each brand–metric pair. Points indicate mean sentiment scores, and error bars show 95\% confidence intervals.
Across brands, sentiment scores exhibit a clear stratification across metrics. Self-Expression consistently receives the highest mean scores, whereas tactile- and audio-related metrics tend to receive lower scores and show greater variability. Several metrics, such as Reliability, show pronounced brand differences with largely non-overlapping confidence intervals, while others, including Ease of Use and Auditory Effects, exhibit more overlapping intervals, indicating more similar evaluations across brands.

\begin{table}[ht]
\caption{Representative evidence-traceability examples showing how localized evidence units are mapped to hierarchical smartphone UX metrics and assigned satisfaction sentiment scores (1--5).}
\label{tab:evidence_traceability_examples}
    \centering
    \small
\begin{tabular}{|c|p{6.8cm}|p{5cm}|c|}
\hline
\rowcolor[HTML]{C0C0C0} 
\textbf{ID} & \textbf{Evidence unit} & \textbf{Metric path (L1 $\rightarrow$ L2 $\rightarrow$ L3 $\rightarrow$ L4)} & \textbf{Score} \\ \hline
E1 & \textit{``Completely unusable, it crashes constantly. I regret buying it.''} &  Functional Experience $\rightarrow$ Reliability $\rightarrow$ Stability $\rightarrow$ Third-Party Software Stability & 1 \\ \hline
E2 & \textit{``Not great overall. The Bluetooth drops sometimes and it is annoying.''} & Functional Experience $\rightarrow$ Reliability $\rightarrow$ Stability $\rightarrow$ Connection Stability & 2 \\ \hline
E3 & \textit{``It is okay. The battery is fine, but the camera is just average.''} & Functional Experience $\rightarrow$ Durability $\rightarrow$ Battery Life $\rightarrow$ Daily Usage Battery Life & 3 \\ \hline
E4 & \textit{``Pretty good in daily use. Photos look nice and the screen is clear.''} & Sensory Experience $\rightarrow$ Visual Appeal $\rightarrow$ Display Quality $\rightarrow$ Base Display Performance & 4 \\ \hline
E5 & \textit{``Absolutely love the design. It fits my style perfectly. Highly recommend.''} & Social Experience $\rightarrow$ Self-Expression $\rightarrow$ Personal Style Expression & 5 \\ \hline
\end{tabular}

\end{table}

To illustrate how the hierarchy supports interpretable drill down beyond the main L2 comparisons, we decompose two high-salience metrics with contrasting valence profiles using an aggregate view pooled across all brands and platforms. We use this pooled summary to highlight stable, commonly discussed sub-metrics that are less sensitive to brand- or platform-specific sparsity, and to provide a concrete example of how an L2 metric can be traced to its constituent L3 and L4 metrics.
Visual Appeal has 13,276 mentions with a mean sentiment score of 3.40 and is primarily composed of  Image Aesthetics (63.3\%) and Display Quality (28.8\%), with smaller contributions from Appearance and Design and Animation Effects. Representative Level 4 metrics include Camera Performance (5,068 mentions) and Base Display Performance (2,140 mentions). In contrast, Reliability has 7,542 mentions with a lower mean sentiment of 2.84 and is dominated by Stability (88.1\%), with smaller contributions from Security and Responsiveness. Representative Level 4 metrics include Connection Stability (2,034 mentions) and Network Stability (1,942 mentions).
This contrast suggests that positive visual impressions are anchored in imaging and display sub-metrics, whereas negative evaluations of reliability are concentrated in stability-related failures. To demonstrate evidence traceability, Table~\ref{tab:evidence_traceability_examples} shows representative localized evidence units with their hierarchical metric-path assignments and sentiment labels.

\subsection{Platform Effects on Salience and Valence Within Brands}
\begin{figure}[ht]
    \centering
    \includegraphics[width=0.8\textwidth]{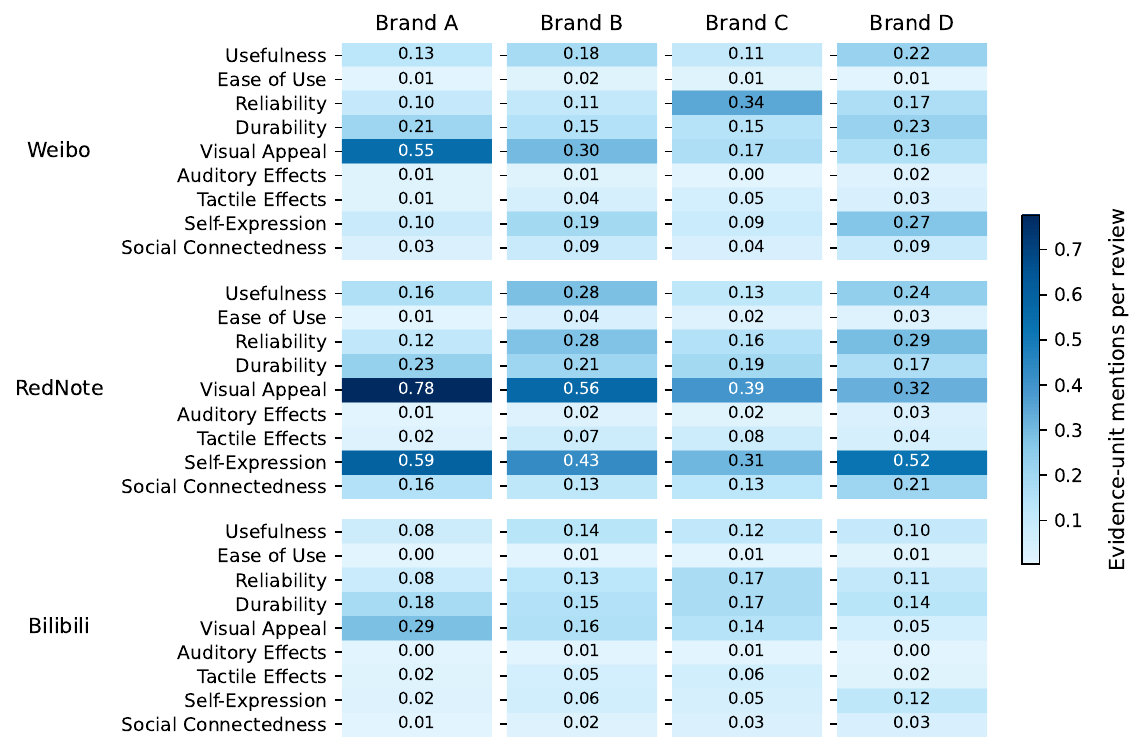}
    \caption{Platform stratified salience of L2 experience metrics within each brand.}
    \label{fig:8}
\end{figure}

\begin{figure}[ht]
    \centering
    \includegraphics[width=0.8\textwidth]{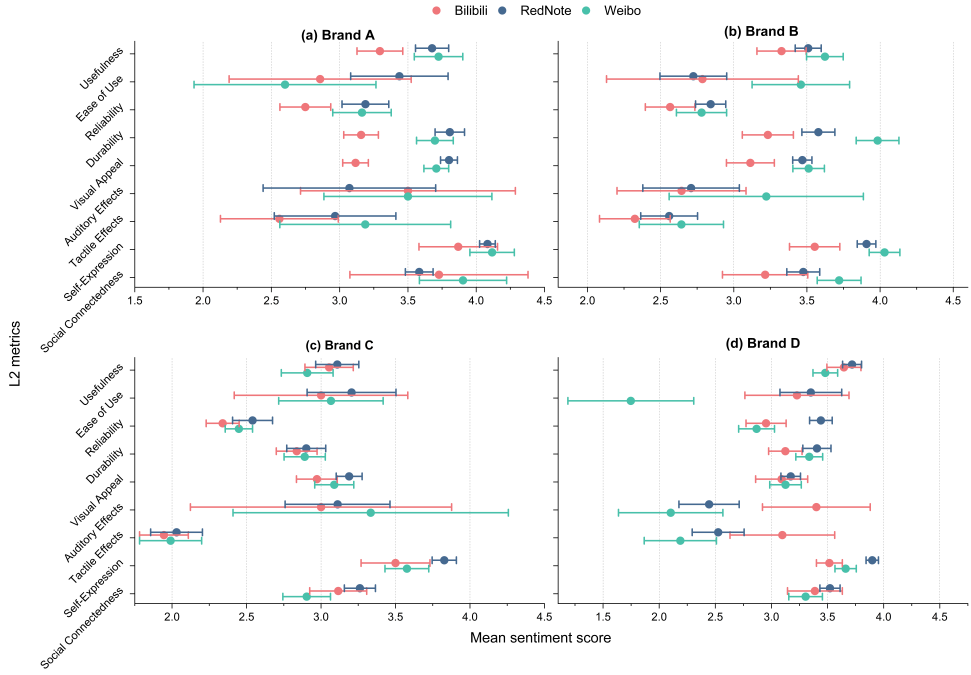}
    \caption{Platform stratified valence of L2 experience metrics within each brand.}
    \label{fig:7}
\end{figure}

This section examines whether platform context is associated with within-brand differences in what users discuss and how they evaluate those experience metrics. We report platform-stratified salience as evidence-unit mentions per review (Figure \ref{fig:8}) and platform-stratified sentiment estimates with 95\% confidence intervals (Figure \ref{fig:7}), allowing us to contrast platform variation in content composition versus evaluative patterns.

Across brands, salience distributions show clear platform stratification, most prominently for Visual Appeal, Self-Expression, and Social Connectedness. RedNote consistently amplifies visually and identity-relevant metrics, with Visual Appeal and Self-Expression exhibiting markedly higher mentions-per-review than on Weibo or Bilibili. In contrast, Bilibili shows lower salience for SX and Social Connectedness, while functional metrics are relatively more prominent for several brands. Meanwhile, some metrics remain low across all platforms (e.g., Ease of Use and Auditory Effects), indicating that they are rarely volunteered in naturalistic discourse regardless of platform context. Overall, these results indicate the platform context is associated with differences in the metric composition of the expressed content.

In contrast to the salience shifts, platform-stratified sentiment patterns are largely stable for metrics with sufficient sample sizes. For most metrics, mean sentiment scores across Bilibili, RedNote, and Weibo show overlapping 95\% confidence intervals, which is consistent with broadly similar sentiment patterns across platforms. High-valence metrics such as Self-Expression remain consistently positive across platforms for all brands. We observe localized platform-level shifts for certain brand–metric pairs. For example, Reliability is often estimated slightly higher on RedNote than on Bilibili within the same brand. In contrast, metrics with smaller sample sizes on some platforms (e.g., Auditory Effects) exhibit wider confidence intervals, indicating less stable platform comparisons. Collectively, these results suggest that within-brand platform effects are generally weaker than the cross-brand differences observed in Figure \ref{fig:6} and platform-stratified differences appear more pronounced in salience than in valence.

\section{Discussion}

\revise{
Smartphones are pervasive personal computing devices whose UX emerges through everyday, multi-context use. This makes smartphone UX difficult to capture through task-based evaluations, aggregate ratings, or laboratory studies alone. Public reviews offer a complementary lens on expressed experience by showing which device experiences become visible, reportable, and contested in public discourse. We discuss how such review-based evidence can support UX measurement and experience characterization for pervasive personal devices.
}

\subsection{Review-Based UX Measurement as Expressed Experience}

Methodologically, our findings suggest that review-based UX measurement is best understood as a way to study how device experience becomes publicly articulated, rather than as a substitute for representative satisfaction surveys. 
\revise{This makes public reviews a form of naturally occurring in-the-wild UX evidence, rather than a controlled record of individual satisfaction.}
Public reviews record experiences that users notice, can attribute to the device, and consider worth formulating for an audience~\cite{whitney-jocelynkouahoInvestigatingPerspectivesExperiences2024,mohammedaldeenEndusersKnowBest2024}. 
This makes review data selective, but the selectivity is analytically meaningful. 
\revise{Rather than treating this selectivity only as a sampling limitation, we use it to examine which device experiences become publicly visible and actionable in everyday discourse.}
It reveals which parts of smartphone experience become reportable in everyday discourse, and which parts remain less visible unless they fail, become socially salient, or are framed through brand-related stance-taking.

This perspective also clarifies how salience and valence should be interpreted. 
Salience should not be equated with importance. Metrics such as photography, battery life, smoothness, and display quality are easier to observe, attribute, and compare across devices. By contrast, learnability, memory management, or social connectedness may be backgrounded, cumulative, or difficult to isolate as discrete product attributes.
Salience comparisons across functional, sensory, and social metrics should be read as differences in expressed visibility. They indicate which experience types are more frequently surfaced in public discourse under our operationalization, rather than which types are intrinsically more important to users. This interpretation is especially important because the pipeline uses evidence-unit formulations suited to different experience types. The unified sentiment model provides a common evaluative scale for extracted evidence units, while valence refers to evaluations of expressed evidence rather than overall device satisfaction.

\subsection{Multi-Layer Smartphone UX and the Uneven Visibility of Experience}

A second implication concerns how smartphone UX is conceptualized. The findings suggest that smartphone UX should be understood as a multi-layer construct whose components differ in how they become visible in public discourse. Functional experience is often expressed through concrete and attributable issues, such as battery life, smoothness, photography, stability, and responsiveness. These qualities are comparatively easy for users to notice, evaluate, and compare across devices. Sensory experience is also highly reportable because smartphones are continuously seen, touched, heard, and carried. Visual appearance, display quality, thermal perception, haptic feedback, and audio quality are not peripheral embellishments. They are part of how a personal device is encountered in everyday use.

This uneven visibility is also reflected in the observed distribution of review evidence. Concrete functional and sensory metrics were more frequently surfaced, while social metrics appeared more selectively and often through identity- or community-related expressions. 
\revise{Social experience tended to become visible through different forms of discourse.}
It is less likely to appear as a simple feature complaint and more likely to be embedded in narratives of identity, taste, interpersonal recognition, or brand-related self-positioning. This makes social experience harder to capture with usability-centered instruments or flat review mining categories. Its value in our framework is therefore not merely that it adds another category to smartphone UX, but that it makes socially situated meanings operationalizable alongside functional and sensory qualities. 
For HCI and Ubicomp, this matters because smartphones are not only computational tools. They are persistent personal artifacts through which users manage practical tasks, sensory encounters, and social positioning at the same time. 
\revise{A hierarchical UX representation therefore allows review-based analysis to move beyond feature-level opinion mining and toward a more interpretable account of how pervasive personal devices are experienced in everyday life}~\cite{akhilmathurMovingMarketResearch2017}.

\subsection{Metric-Level Diagnosis for Design and Experience Governance}

\revise{
We use experience governance to refer to the ongoing monitoring, interpretation, and prioritization of user-visible experience issues across product iterations. In this context, the value of the framework lies in separating experience quality from experience visibility, rather than simply identifying positive and negative evaluations. For UbiComp research, this distinction highlights that large-scale UX measurement should examine which experiences become publicly observable, which remain under-articulated, and how these visibility patterns shape interpretation.
}
Consistent with prior work that uses reviews for fine-grained comparison beyond overall ratings~\cite{yuanchunliMiningUserReviews2017}, this metric-level view avoids reducing review data to aggregate sentiment alone.
This shifts review analysis from ranking products or brands to examining how particular experience qualities become visible, contested, or taken for granted in public discourse. 
A metric may receive negative evaluations because it is genuinely problematic, because it is highly visible and easy to complain about, or because platform norms encourage users to discuss it. Conversely, a metric may appear unimportant simply because it is difficult to articulate, rarely foregrounded, or taken for granted when it works well. Metric-level diagnosis therefore helps design teams avoid treating review volume as a direct proxy for design importance.

The salience--valence lens can be used as an interpretive tool for experience governance. High-salience and low-valence metrics indicate visible frictions that have already become public experience problems. High-salience and high-valence metrics indicate visible strengths that shape product differentiation. \revise{When sufficient evidence is available, low-salience and low-valence metrics should not be dismissed, because they may reflect silent risks or under-articulated breakdowns that require interviews, diary studies, or targeted surveys.} Low-salience and high-valence metrics may suggest design strengths that users value when encountered, but that remain weakly visible in public discussion. \revise{In this sense, the framework supports prioritization, follow-up user research, and decisions about which visible frictions require design attention and where design value needs to be more clearly surfaced in use.}

\subsection{\revise{Transferability and Contextual Boundaries}}
\label{sec:discussion_generalizability}

\revise{
Although the empirical setting of this study is Chinese social media, the contribution should be read primarily as methodological. 
We distinguish the transferability of the approach from the generalizability of the empirical patterns. 
The panel review, end-to-end validation, and four-brand application support the framework's task-specific clarity, operationalization, and analytical utility in the studied setting. 
They do not establish universal validity across review environments. 
The transferable element is the use of a hierarchical UX representation to organize naturally occurring user feedback into interpretable evidence about smartphone experience. 
This approach provides a basis for adapting review-based UX measurement to other review environments, languages, markets, and pervasive personal devices where users publicly describe situated experiences.
}

\revise{
The empirical patterns reported in this paper should not be interpreted as universal properties of smartphone UX. 
Metric salience, valence, and brand-level comparisons are shaped by the studied platforms, selected flagship devices, market context, and collection period. 
Platform norms and audience composition may affect which experiences become publicly visible. 
Cultural and market contexts may affect how users discuss service quality, ecosystem dependence, privacy, brand identity, and social meanings of smartphone use. 
Our findings are therefore best understood as context-specific evidence of expressed smartphone UX rather than population-level estimates across all smartphone users or markets.
}

\revise{
The linguistic context is another boundary. 
Chinese social media comments often include colloquial expressions, abbreviations, implicit evaluations, exaggeration, sarcasm, and platform-specific slang. 
These features may affect evidence-unit extraction, metric mapping, and satisfaction interpretation. 
Applying the framework to other languages or platforms would therefore require context-specific adaptation and revalidation, including the metric definitions, annotation guidelines, and the extraction, classification, and valence models. 
The concrete L3/L4 metrics and their distributions should therefore be treated as adaptable operational categories rather than as a fixed universal taxonomy.
}

\section{Limitations}

Our findings should be interpreted with three limitations in mind. 
First, online reviews are selective expressions of experience and are subject to selection, extremity, and stance-related biases. 
Less vocal users and less intense experiences may be underrepresented, and some metrics may remain weakly expressed unless breakdowns occur. 
Public review discourse may also contain polarized brand-related stance-taking, which can amplify sentiment and blur the boundary between product experience, brand identity, and community-based positioning. 
Although our evidence-unit extraction reduces the influence of review-level polarization by focusing on localized experience expressions, it cannot fully separate experience evidence from brand-related stance-taking. 
\revise{Salience and valence should therefore be interpreted as properties of expressed review evidence rather than estimates of population-level prevalence or overall satisfaction.}
\revise{This boundary is especially relevant for fine-grained L3/L4 outputs. Because naturally occurring reviews have uneven class distributions, low-support or context-dependent metrics should be interpreted as diagnostic drill-down evidence rather than uniformly stable aggregate measurement endpoints. Accordingly, our main corpus-level comparisons rely on Level 2, while L3/L4 results are used for drill-down interpretation when sufficient evidence is available.
}

Second, our empirical analysis is grounded in three major Chinese social media platforms and focuses on flagship smartphones. 
Platform norms, audience composition, recommendation mechanisms, market structure, product segment, and linguistic conventions may affect what users choose to mention and how they frame evaluations. 
\revise{
Chinese social media discourse may include colloquial expressions, abbreviations, implicit evaluations, rhetorical exaggeration, sarcasm, and platform-specific slang, which may influence evidence-unit extraction, metric mapping, and sentiment interpretation. 
}
The resulting metric distributions, salience patterns, sentiment scores, brand-level findings, and trained classifiers should therefore not be assumed to transfer directly to other languages, platforms, device segments, or cultural settings. 
\revise{
The resulting metric distributions, salience patterns, valence scores, brand-level findings, and trained classifiers should therefore not be assumed to transfer directly to other languages, platforms, device segments, or cultural settings without further adaptation and validation.
}

Third, review-based sentiment should not be treated as a direct proxy for long-term use, adoption, retention, or market performance. 
Reviewers and buyers may not fully overlap, and behavior is shaped by pricing, availability, ecosystem constraints, channel effects, and changes over time. 
Review-based signals are therefore best used as diagnostic evidence of expressed experience. They should be complemented by surveys, interviews, diary studies, experience sampling, or longitudinal data~\cite{danielhintzeLargescaleLongtermAnalysis2017} when the goal is to estimate population-level UX prevalence, behavioral outcomes, or causal effects of design changes.

\section{Conclusion}

This paper addressed the need for scalable and interpretable measurement of smartphone user experience from naturalistic user discourse. We developed a theory-grounded hierarchical measurement framework with consistent construct boundaries across metric levels, and implemented an evidence-traceable computational pipeline that links localized review evidence to hierarchical UX metrics and unified satisfaction scores.
Using approximately 20,000 Chinese social media reviews spanning four major brands and three platforms, we characterized smartphone UX at the Level-2 metric level by separating what users discuss from how they evaluate it. The analysis further showed how platform-stratified results can contextualize within-brand variation across review contexts.
Together, the framework, pipeline, and empirical findings demonstrate how hierarchical UX metrics can support comparable reporting and metric-level diagnosis beyond usability-only measures or single-score satisfaction summaries. 
More broadly, while our empirical analysis is situated in Chinese social media, this work highlights the value of transforming public user discourse into structured UX evidence, supporting a richer understanding of mobile experience as it is expressed and evaluated in everyday digital life.

We would like to express our sincere gratitude to the domain experts who contributed to the annotation and validation of our framework. 
During the preparation of this work, the authors used generative AI tools (e.g., ChatGPT) in two limited ways: 
(1) to assist with the structured extraction of user comments in the methodology by using GPT-4o-mini (in Section~\ref{4.4}), and 
(2) to provide language polishing and stylistic refinement of the manuscript. 
After using these tools, the authors reviewed and edited the content as needed. 
All research ideas, analyses, and interpretations are the authors’ own, and the authors take full responsibility for the final content of this publication.

\bibliographystyle{ACM-Reference-Format}
\bibliography{metal-model-base}

@article{akhilmathurMovingMarketResearch2017,
  title = {Moving beyond Market Research: {{Demystifying}} Smartphone User Behavior in {{India}}},
  shorttitle = {Moving {{Beyond Market Research}}},
  author = {{Akhil Mathur} and {Lakshmi Manasa Kalanadhabhatta} and {Rahul Majethia} and {Fahim Kawsar}},
  year = 2017,
  journal = {Proceedings of the ACM on Interactive, Mobile, Wearable and Ubiquitous Technologies},
  volume = {1},
  number = {3},
  pages = {1--27},
  doi = {10.1145/3130947},
  url = {https://dl.acm.org/doi/10.1145/3130947}
}

@techreport{alibabaclouduxteamEaseUseMetric2020,
  type = {{Group Standard}},
  title = {{Ease of Use Metric for Cloud Product}},
  author = {{Alibaba Cloud UX Team} and {China Industrial Design Association}},
  year = 2020,
  institution = {Alibaba Cloud Computing Co., Ltd.},
  url = {https://www.ttbz.org.cn/upload/file/20191207/6371128187942843383935807.pdf}
}

@misc{anantkhandelwalLargeScaleGenerative2023,
  title = {Large Scale Generative Multimodal Attribute Extraction for {{E-commerce}} Attributes},
  author = {{Anant Khandelwal} and {Happy Mittal} and {Shreyas Sunil Kulkarni} and {Deepak Gupta}},
  year = 2023,
  number = {arXiv:2306.00379},
  eprint = {2306.00379},
  primaryclass = {cs},
  publisher = {arXiv},
  doi = {10.48550/arXiv.2306.00379},
  url = {http://arxiv.org/abs/2306.00379},
  archiveprefix = {arXiv}
}

@inproceedings{andrerusliLeveragingLLMsAttribute2024,
  title = {Leveraging {{LLMs}} for Attribute Definition and Value Extraction from User-Generated Texts on a {{C2C E-commerce}} Platform},
  booktitle = {Proceedings of the 2024 8th {{International Conference}} on {{Natural Language Processing}} and {{Information Retrieval}}},
  author = {{Andre Rusli} and {Yuji Oshima} and {Yuki Yada}},
  year = 2024,
  pages = {186--191},
  publisher = {ACM},
  address = {Okayama Japan},
  doi = {10.1145/3711542.3711599},
  url = {https://dl.acm.org/doi/10.1145/3711542.3711599}
}

@inproceedings{anselblumeGenerativeModelsProduct2023,
  title = {Generative Models for Product Attribute Extraction},
  booktitle = {Proceedings of the 2023 {{Conference}} on {{Empirical Methods}} in {{Natural Language Processing}}: {{Industry Track}}},
  author = {{Ansel Blume} and {Nasser Zalmout} and {Heng Ji} and {Xian Li}},
  year = 2023,
  publisher = {Association for Computational Linguistics},
  address = {Singapore},
  doi = {10.18653/v1/2023.emnlp-industry.55},
  url = {https://aclanthology.org/2023.emnlp-industry.55}
}

@article{anttioulasvirtaHabitsMakeSmartphone2012,
  title = {Habits Make Smartphone Use More Pervasive},
  author = {{Antti Oulasvirta} and {Tye Rattenbury} and {Lingyi Ma} and {Eeva Raita}},
  year = 2012,
  journal = {Personal and Ubiquitous Computing},
  volume = {16},
  number = {1},
  pages = {105--114},
  doi = {10.1007/s00779-011-0412-2},
  url = {http://link.springer.com/10.1007/s00779-011-0412-2}
}

@article{argahanantoApplicationTextMining2016,
  title = {Application of Text Mining to Extract Hotel Attributes and Construct Perceptual Map of Five Star Hotels from Online Review: {{Study}} of Jakarta and Singapore Five-Star Hotels},
  shorttitle = {Application of {{Text Mining}} to {{Extract Hotel Attributes}} and {{Construct Perceptual Map}} of {{Five Star Hotels}} from {{Online Review}}},
  author = {{Arga Hananto}},
  year = 2016,
  journal = {ASEAN Marketing Journal},
  volume = {7},
  number = {2},
  publisher = {Universitas Indonesia},
  doi = {10.21002/amj.v7i2.5262},
  url = {http://journal.ui.ac.id/index.php/amj/article/view/5262}
}

@techreport{baidu-ernie-teamERNIE45Technical2025,
  title = {{{ERNIE}} 4.5 Technical Report},
  author = {{Baidu-ERNIE-Team}},
  year = 2025,
  url = {https://github.com/PaddlePaddle/ERNIE}
}

@book{berndh.schmittExperientialMarketingHow2000,
  title = {Experiential Marketing: {{How}} to Get Customers to Sense, Feel, Think, Act, Relate},
  shorttitle = {Experiential {{Marketing}}},
  author = {{Bernd H. Schmitt}},
  year = 2000,
  publisher = {Free Press},
  address = {Riverside},
  isbn = {978-0-7432-1951-8}
}

@inproceedings{bhaveshkukrejaSentimentMultifacetedReview2024,
  title = {Beyond Sentiment: {{A}} Multifaceted Review Scoring System for Enhanced Customer Feedback Analysis},
  shorttitle = {Beyond {{Sentiment}}},
  booktitle = {Proceedings of the 8th {{International Conference}} on {{Data Science}} and {{Management}} of {{Data}} (12th {{ACM IKDD CODS}} and 30th {{COMAD}})},
  author = {{Bhavesh Kukreja} and {Aritra Ghosh Dastidar} and {Radhika Mundra} and {Kartikey Singh} and {Javaid Nabi}},
  year = 2024,
  pages = {244--251},
  publisher = {ACM},
  address = {Jodhpur India},
  doi = {10.1145/3703323.3703729},
  url = {https://dl.acm.org/doi/10.1145/3703323.3703729}
}

@book{bingliuSentimentAnalysisOpinion2012,
  title = {Sentiment Analysis and Opinion Mining},
  author = {{Bing Liu}},
  year = 2012,
  series = {Synthesis {{Lectures}} on {{Human Language Technologies}}},
  publisher = {Springer International Publishing},
  address = {Cham},
  doi = {10.1007/978-3-031-02145-9},
  url = {https://link.springer.com/10.1007/978-3-031-02145-9},
  isbn = {978-3-031-01017-0 978-3-031-02145-9}
}

@inproceedings{chaoxupangGuidelineLearningIncontext2023,
  title = {Guideline Learning for In-Context Information Extraction},
  booktitle = {Proceedings of the 2023 {{Conference}} on {{Empirical Methods}} in {{Natural Language Processing}}},
  author = {{Chaoxu Pang} and {Yixuan Cao} and {Qiang Ding} and {Ping Luo}},
  year = 2023,
  pages = {15372--15389},
  publisher = {Association for Computational Linguistics},
  address = {Singapore},
  doi = {10.18653/v1/2023.emnlp-main.950},
  url = {https://aclanthology.org/2023.emnlp-main.950}
}

@incollection{chengjunzhouPaperResearchMobile2024,
  title = {Paper Research on Mobile Phone Brand Image Design Based on Sensory Marketing Theory--Taking Apple as an Example},
  booktitle = {Man-{{Machine-Environment System Engineering}}},
  author = {{Chengjun Zhou} and {Chunxing Hao} and {Yuhuan Chi} and {Mengtian Sun}},
  editor = {{Shengzhao Long} and {Balbir S. Dhillon} and {Long Ye}},
  year = 2024,
  volume = {1256},
  pages = {407--417},
  publisher = {Springer Nature Singapore},
  address = {Singapore},
  doi = {10.1007/978-981-97-7139-4_56},
  url = {https://link.springer.com/10.1007/978-981-97-7139-4_56},
  isbn = {978-981-97-7138-7 978-981-97-7139-4}
}

@article{chengyangOnlineUserReview2021,
  title = {Online User Review Analysis for Product Evaluation and Improvement},
  author = {{Cheng Yang} and {Lingang Wu} and {Kun Tan} and {Chunyang Yu} and {Yuliang Zhou} and {Ye Tao} and {Yu Song}},
  year = 2021,
  journal = {Journal of Theoretical and Applied Electronic Commerce Research},
  volume = {16},
  number = {5},
  pages = {1598--1611},
  doi = {10.3390/jtaer16050090},
  url = {https://www.mdpi.com/0718-1876/16/5/90}
}

@article{chituokoliDelphiMethodResearch2004,
  title = {The Delphi Method as a Research Tool: {{An}} Example, Design Considerations and Applications},
  shorttitle = {The {{Delphi}} Method as a Research Tool},
  author = {{Chitu Okoli} and {Suzanne D. Pawlowski}},
  year = 2004,
  journal = {Information \& Management},
  volume = {42},
  number = {1},
  pages = {15--29},
  doi = {10.1016/j.im.2003.11.002},
  url = {https://linkinghub.elsevier.com/retrieve/pii/S0378720603001794}
}

@misc{datareportalDigitalWorld2026,
  title = {Digital around the World},
  author = {{DataReportal}},
  year = 2026,
  journal = {DataReportal -- Global Digital Insights},
  url = {https://datareportal.com/global-digital-overview}
}

@article{derongxuLargeLanguageModels2024,
  title = {Large Language Models for Generative Information Extraction: {{A}} Survey},
  shorttitle = {Large Language Models for Generative Information Extraction},
  author = {{Derong Xu} and {Wei Chen} and {Wenjun Peng} and {Chao Zhang} and {Tong Xu} and {Xiangyu Zhao} and {Xian Wu} and {Yefeng Zheng} and {Yang Wang} and {Enhong Chen}},
  year = 2024,
  journal = {Frontiers of Computer Science},
  volume = {18},
  number = {6},
  pages = {186357},
  doi = {10.1007/s11704-024-40555-y},
  url = {https://link.springer.com/10.1007/s11704-024-40555-y}
}

@inproceedings{divyamsobtiDomainspecificAspectExtraction2025,
  title = {Domain-Specific Aspect Extraction for Product Design},
  booktitle = {Companion {{Proceedings}} of the {{ACM}} on {{Web Conference}} 2025},
  author = {{Divyam Sobti} and {Mahima Agumbe Suresh} and {Cristina Tortora} and {Vimal Viswanathan}},
  year = 2025,
  pages = {2754--2763},
  publisher = {ACM},
  address = {Sydney NSW Australia},
  doi = {10.1145/3701716.3717851},
  url = {https://dl.acm.org/doi/10.1145/3701716.3717851}
}

@book{donalda.normanEmotionalDesignWhy2004,
  title = {Emotional Design: {{Why}} We Love (or Hate) Everyday Things},
  shorttitle = {Emotional Design},
  author = {{Donald A. Norman}},
  year = 2004,
  publisher = {Basic Books},
  address = {New York},
  isbn = {978-0-465-05135-9},
  lccn = {BF531 .N67 2004}
}

@article{ehsanmortazaviExploringLandscapeUX2024,
  title = {Exploring the Landscape of {{UX}} Subjective Evaluation Tools and {{UX}} Dimensions: {{A}} Systematic Literature Review (2010--2021)},
  shorttitle = {Exploring the {{Landscape}} of {{UX Subjective Evaluation Tools}} and {{UX Dimensions}}},
  author = {{Ehsan Mortazavi} and {Philippe Doyon-Poulin} and {Daniel Imbeau} and {Mitra Taraghi} and {Jean-Marc Robert}},
  year = 2024,
  journal = {Interacting with Computers},
  volume = {36},
  number = {4},
  pages = {255--278},
  doi = {10.1093/iwc/iwae017},
  url = {https://academic.oup.com/iwc/article/36/4/255/7685409}
}

@article{erkanbayraktarMeasuringEfficiencyCustomer2012a,
  title = {Measuring the Efficiency of Customer Satisfaction and Loyalty for Mobile Phone Brands with {{DEA}}},
  author = {{Erkan Bayraktar} and {Ekrem Tatoglu} and {Ali Turkyilmaz} and {Dursun Delen} and {Selim Zaim}},
  year = 2012,
  journal = {Expert Systems with Applications},
  volume = {39},
  number = {1},
  pages = {99--106},
  doi = {10.1016/j.eswa.2011.06.041},
  url = {https://linkinghub.elsevier.com/retrieve/pii/S0957417411009419}
}

@misc{fabioduarteTimeSpentUsing2025,
  title = {Time Spent Using Smartphones (2025 Statistics)},
  author = {{Fabio Duarte}},
  year = 2025,
  journal = {Exploding Topics},
  url = {https://explodingtopics.com/blog/smartphone-usage-stats}
}

@article{fangminchengUserExperienceEvaluation2021,
  title = {User Experience Evaluation Method Based on Online Product Reviews},
  author = {{Fangmin Cheng} and {Suihuai Yu} and {Shengfeng Qin} and {Jianjie Chu} and {Jian Chen}},
  year = 2021,
  journal = {Journal of Intelligent \& Fuzzy Systems},
  volume = {41},
  number = {1},
  pages = {1791--1805},
  doi = {10.3233/JIFS-210564},
  url = {https://journals.sagepub.com/doi/full/10.3233/JIFS-210564}
}

@article{fenghuMappingHotelBrand2020,
  title = {Mapping Hotel Brand Positioning and Competitive Landscapes by Text-Mining User-Generated Content},
  author = {{Feng Hu} and {Rohit H. Trivedi}},
  year = 2020,
  journal = {International Journal of Hospitality Management},
  volume = {84},
  pages = {102317},
  doi = {10.1016/j.ijhm.2019.102317},
  url = {https://linkinghub.elsevier.com/retrieve/pii/S0278431918309496}
}

@article{fredd.davisPerceivedUsefulnessPerceived1989,
  title = {Perceived Usefulness, Perceived Ease of Use, and User Acceptance of Information Technology},
  author = {{Fred D. Davis}},
  year = 1989,
  journal = {MIS Quarterly},
  volume = {13},
  number = {3},
  eprint = {249008},
  eprinttype = {jstor},
  pages = {319},
  publisher = {JSTOR},
  doi = {10.2307/249008},
  url = {https://www.jstor.org/stable/249008?origin=crossref}
}

@article{gangkouCrossplatformMarketStructure2021,
  title = {A Cross-Platform Market Structure Analysis Method Using Online Product Reviews},
  author = {{Gang Kou} and {Pei Yang} and {Yi Peng} and {Hui Xiao} and {Feng Xiao} and {Yang Chen} and {Fawaz E. Alsaadi}},
  year = 2021,
  journal = {Technological and Economic Development of Economy},
  volume = {27},
  number = {5},
  pages = {992--1018},
  doi = {10.3846/tede.2021.12005},
  url = {https://journals.vilniustech.lt/index.php/TEDE/article/view/12005}
}

@article{haiyunpengReviewSentimentAnalysis2017,
  title = {A Review of Sentiment Analysis Research in {{Chinese}} Language},
  author = {{Haiyun Peng} and {Erik Cambria} and {Amir Hussain}},
  year = 2017,
  journal = {Cognitive Computation},
  volume = {9},
  number = {4},
  pages = {423--435},
  doi = {10.1007/s12559-017-9470-8},
  url = {http://link.springer.com/10.1007/s12559-017-9470-8}
}

@article{heUnderstandingConsumersMulticompeting2023,
  title = {Understanding the Consumers' Multi-Competing Brand Community Engagement: {{A}} Mix Method Approach},
  shorttitle = {Understanding the Consumers' Multi-Competing Brand Community Engagement},
  author = {He, Kai and Liao, Junyun and Li, Fengyan and Sun, Hongguang},
  year = 2023,
  journal = {Frontiers in Psychology},
  volume = {13},
  pages = {1088619},
  doi = {10.3389/fpsyg.2022.1088619},
  url = {https://www.frontiersin.org/articles/10.3389/fpsyg.2022.1088619/full}
}

@article{huizhangProductInnovationBased2018,
  title = {Product Innovation Based on Online Review Data Mining: {{A}} Case Study of Huawei Phones},
  shorttitle = {Product Innovation Based on Online Review Data Mining},
  author = {{Hui Zhang} and {Huguang Rao} and {Junzheng Feng}},
  year = 2018,
  journal = {Electronic Commerce Research},
  volume = {18},
  number = {1},
  pages = {3--22},
  doi = {10.1007/s10660-017-9279-2},
  url = {http://link.springer.com/10.1007/s10660-017-9279-2}
}

@misc{idcSmartphoneMarketShare2026,
  title = {Smartphone Market Share},
  author = {{IDC}},
  year = 2026,
  journal = {IDC Tracker},
  url = {https://www.idc.com/promo/smartphone-market-share/market-share/}
}

@techreport{internationalorganizationforstandardizationErgonomicRequirementsOffice1998,
  type = {Standard},
  title = {Ergonomic Requirements for Office Work with Visual Display Terminals ({{VDTs}}) - Part 11: {{Guidance}} on Usability},
  author = {{International Organization for Standardization}},
  year = 1998,
  number = {ISO 9241-11:1998},
  institution = {International Organization for Standardization}
}

@techreport{internationalorganizationforstandardizationErgonomicsHumansystemInteraction2019,
  type = {Standard},
  title = {Ergonomics of Human-System Interaction - Part 210: {{Human-centred}} Design for Interactive Systems},
  author = {{International Organization for Standardization}},
  year = 2019,
  number = {ISO 9241-210:2019},
  address = {Geneva, Switzerland},
  institution = {International Organization for Standardization},
  url = {https://www.iso.org/standard/77520.html}
}

@book{jakobnielsenUsabilityEngineering1993,
  title = {Usability Engineering},
  author = {{Jakob Nielsen}},
  year = 1993,
  publisher = {Academic Press},
  address = {Boston},
  isbn = {978-0-12-518406-9},
  lccn = {005.1}
}

@article{janvanderlindenUserExperienceUX2025,
  title = {User Experience ({{UX}}) with Mobile Devices: {{A}} Comprehensive Model to Demonstrate the Relative Importance of Instrumental, Non-Instrumental, and Emotional Components on User Satisfaction},
  shorttitle = {User {{Experience}} ({{UX}}) with {{Mobile Devices}}},
  author = {{Jan Van Der Linden} and {Catherine Hellemans} and {Franck Amadieu} and {Emilie Vayre} and {C\'ecile Van De Leemput}},
  year = 2025,
  journal = {International Journal of Human--Computer Interaction},
  volume = {41},
  number = {8},
  pages = {4516--4527},
  doi = {10.1080/10447318.2024.2352210},
  url = {https://www.tandfonline.com/doi/full/10.1080/10447318.2024.2352210}
}

@misc{jeffsaurophdSUPRqmQuestionnaireMeasure2017,
  title = {{{SUPR-qm}}: {{A}} Questionnaire to Measure the Mobile App User Experience - {{JUX}}},
  shorttitle = {{{SUPR-Qm}}},
  author = {{Jeff Sauro, PhD} and {Paree Zarolia}},
  year = 2017,
  journal = {JUX - The Journal of User Experience},
  url = {https://uxpajournal.org/supr-qm-measure-mobile-ux/}
}

@article{jeongyunheoFrameworkEvaluatingUsability2009,
  title = {A Framework for Evaluating the Usability of Mobile Phones Based on Multi-Level, Hierarchical Model of Usability Factors},
  author = {{Jeongyun Heo} and {Dong-Han Ham} and {Sanghyun Park} and {Chiwon Song} and {Wan Chul Yoon}},
  year = 2009,
  journal = {Interacting with Computers},
  volume = {21},
  number = {4},
  pages = {263--275},
  doi = {10.1016/j.intcom.2009.05.006},
  url = {https://academic.oup.com/iwc/article-lookup/doi/10.1016/j.intcom.2009.05.006}
}

@inproceedings{jiatanFrameworkSoftwareUsability2013,
  title = {A Framework for Software Usability and User Experience Measurement in Mobile Industry},
  booktitle = {2013 {{Joint Conference}} of the 23rd {{International Workshop}} on {{Software Measurement}} and the 8th {{International Conference}} on {{Software Process}} and {{Product Measurement}}},
  author = {{Jia Tan} and {Kari Ronkko} and {Cigdem Gencel}},
  year = 2013,
  pages = {156--164},
  publisher = {IEEE},
  address = {Ankara, Turkey},
  doi = {10.1109/IWSM-Mensura.2013.31},
  url = {http://ieeexplore.ieee.org/document/6693235/},
  isbn = {978-0-7695-5078-7}
}

@article{jiegaoExploringSmartphoneUser2024,
  title = {Exploring Smartphone User Interface Experience-Sharing Behavior: {{Design}} Perception and Motivation-Driven Mechanisms through the {{SOR}} Model},
  shorttitle = {Exploring {{Smartphone User Interface Experience-Sharing Behavior}}},
  author = {{Jie Gao} and {Wenjing Jia} and {Jun Yin}},
  year = 2024,
  journal = {Sustainability},
  volume = {16},
  number = {15},
  pages = {6670},
  doi = {10.3390/su16156670},
  url = {https://www.mdpi.com/2071-1050/16/15/6670}
}

@article{jielouEffectsMobileIdentity2022,
  title = {Effects of Mobile Identity on Smartphone Symbolic Use: {{An}} Attachment Theory Perspective},
  shorttitle = {Effects of {{Mobile Identity}} on {{Smartphone Symbolic Use}}},
  author = {{Jie Lou} and {Nianlong Han} and {Dong Wang} and {Xi Pei}},
  year = 2022,
  journal = {International Journal of Environmental Research and Public Health},
  volume = {19},
  number = {21},
  pages = {14036},
  doi = {10.3390/ijerph192114036},
  url = {https://www.mdpi.com/1660-4601/19/21/14036}
}

@article{jihuacaoOnlineReviewsSentiment2023,
  title = {Online Reviews Sentiment Analysis and Product Feature Improvement with Deep Learning},
  author = {{Jihua Cao} and {Jie Li} and {Miao Yin} and {Yunfeng Wang}},
  year = 2023,
  journal = {ACM Transactions on Asian and Low-Resource Language Information Processing},
  volume = {22},
  number = {8},
  pages = {1--17},
  doi = {10.1145/3522575},
  url = {https://dl.acm.org/doi/10.1145/3522575}
}

@inproceedings{jodiforlizziUnderstandingExperienceInteractive2004,
  title = {Understanding Experience in Interactive Systems},
  booktitle = {Proceedings of the 5th Conference on {{Designing}} Interactive Systems: Processes, Practices, Methods, and Techniques},
  author = {{Jodi Forlizzi} and {Katja Battarbee}},
  year = 2004,
  pages = {261--268},
  publisher = {ACM},
  address = {Cambridge MA USA},
  doi = {10.1145/1013115.1013152},
  url = {https://dl.acm.org/doi/10.1145/1013115.1013152},
  isbn = {978-1-58113-787-3}
}

@article{johndagdelenStructuredInformationExtraction2024,
  title = {Structured Information Extraction from Scientific Text with Large Language Models},
  author = {{John Dagdelen} and {Alexander Dunn} and {Sanghoon Lee} and {Nicholas Walker} and {Andrew S. Rosen} and {Gerbrand Ceder} and {Kristin A. Persson} and {Anubhav Jain}},
  year = 2024,
  journal = {Nature Communications},
  volume = {15},
  number = {1},
  pages = {1418},
  doi = {10.1038/s41467-024-45563-x},
  url = {https://www.nature.com/articles/s41467-024-45563-x}
}

@article{johnmccarthyTechnologyExperience2004,
  title = {Technology as Experience},
  author = {{John McCarthy} and {Peter Wright}},
  year = 2004,
  journal = {Interactions},
  volume = {11},
  number = {5},
  pages = {42--43},
  doi = {10.1145/1015530.1015549},
  url = {https://dl.acm.org/doi/10.1145/1015530.1015549}
}

@article{jonahbergerWordMouthInterpersonal2014,
  title = {Word of Mouth and Interpersonal Communication: {{A}} Review and Directions for Future Research},
  shorttitle = {Word of Mouth and Interpersonal Communication},
  author = {{Jonah Berger}},
  year = 2014,
  journal = {Journal of Consumer Psychology},
  volume = {24},
  number = {4},
  pages = {586--607},
  doi = {10.1016/j.jcps.2014.05.002},
  url = {https://myscp.onlinelibrary.wiley.com/doi/10.1016/j.jcps.2014.05.002}
}

@misc{joshhowarthHowManyPeople2021,
  title = {How Many People Own Smartphones? (2025-2029)},
  shorttitle = {How {{Many People Own Smartphones}}?},
  author = {{Josh Howarth}},
  year = 2021,
  journal = {Exploding Topics},
  url = {https://explodingtopics.com/blog/smartphone-stats}
}

@article{kevinlanekellerConceptualizingMeasuringManaging1993,
  title = {Conceptualizing, Measuring, and Managing Customer-Based Brand Equity},
  author = {{Kevin Lane Keller}},
  year = 1993,
  journal = {Journal of Marketing},
  volume = {57},
  number = {1},
  pages = {1--22},
  doi = {10.1177/002224299305700101},
  url = {https://journals.sagepub.com/doi/10.1177/002224299305700101}
}

@book{kevinlanekellerStrategicBrandManagement2020,
  title = {Strategic Brand Management: Building, Measuring, and Managing Brand Equity},
  shorttitle = {Strategic Brand Management},
  author = {{Kevin Lane Keller} and {Vanitha Swaminathan}},
  year = 2020,
  edition = {Fifth edition},
  publisher = {Pearson},
  address = {Hoboken, NJ},
  isbn = {978-0-13-489249-8},
  lccn = {HD69.B7 K45 2020}
}

@article{kexuanniuEventawareSarcasmDetection2025,
  title = {Event-Aware Sarcasm Detection in {{Chinese}} Social Media Using Multi-Head Attention and Contrastive Learning},
  author = {{Kexuan Niu} and {Xiameng Si} and {Xiaojie Qi} and {Haiyan Kang}},
  year = 2025,
  journal = {Computers, Materials \& Continua},
  volume = {85},
  number = {1},
  pages = {2051--2070},
  doi = {10.32604/cmc.2025.065377},
  url = {https://www.techscience.com/cmc/v85n1/63518}
}

@article{lalehdavoodiAutomatingCustomerFeedback2026,
  title = {Automating Customer Feedback Analysis in {{E-commerce}}: {{A}} Multi-{{Model}} Approach},
  shorttitle = {Automating Customer Feedback Analysis in {{E-commerce}}},
  author = {{Laleh Davoodi} and {J\'ozsef Mezei} and {Shahrokh Nikou} and {Leonardo Espinosa-Leal}},
  year = 2026,
  journal = {Expert Systems with Applications},
  volume = {306},
  pages = {130865},
  doi = {10.1016/j.eswa.2025.130865},
  url = {https://linkinghub.elsevier.com/retrieve/pii/S095741742504480X}
}

@article{leehumphreysMobileSocialMedia2013,
  title = {Mobile Social Media: {{Future}} Challenges and Opportunities},
  shorttitle = {Mobile Social Media},
  author = {{Lee Humphreys}},
  year = 2013,
  journal = {Mobile Media \& Communication},
  volume = {1},
  number = {1},
  pages = {20--25},
  doi = {10.1177/2050157912459499},
  url = {https://journals.sagepub.com/doi/10.1177/2050157912459499}
}

@article{leihuangSurveyHallucinationLarge2025,
  title = {A Survey on Hallucination in Large Language Models: {{Principles}}, Taxonomy, Challenges, and Open Questions},
  shorttitle = {A {{Survey}} on {{Hallucination}} in {{Large Language Models}}},
  author = {{Lei Huang} and {Weijiang Yu} and {Weitao Ma} and {Weihong Zhong} and {Zhangyin Feng} and {Haotian Wang} and {Qianglong Chen} and {Weihua Peng} and {Xiaocheng Feng} and {Bing Qin} and {Ting Liu}},
  year = 2025,
  journal = {ACM Transactions on Information Systems},
  volume = {43},
  number = {2},
  pages = {1--55},
  doi = {10.1145/3703155},
  url = {https://dl.acm.org/doi/10.1145/3703155}
}

@article{lilishiProductFeatureExtraction2023,
  title = {Product Feature Extraction from {{Chinese}} Online Reviews: {{Application}} to Product Improvement},
  shorttitle = {Product Feature Extraction from {{Chinese}} Online Reviews},
  author = {{Lili Shi} and {Jun Lin} and {Guoquan Liu}},
  year = 2023,
  journal = {RAIRO - Operations Research},
  volume = {57},
  number = {3},
  pages = {1125--1147},
  doi = {10.1051/ro/2023046},
  url = {https://www.rairo-ro.org/10.1051/ro/2023046}
}

@article{liuguohaPerceivedImpactProduct2025,
  title = {Perceived Impact of Product Innovation on Brand Loyalty: {{A}} Study of Smartphone Consumers in {{China}}},
  shorttitle = {Perceived {{Impact}} of {{Product Innovation}} on {{Brand Loyalty}}},
  author = {{Liu Guoha} and {Rovena Dellova}},
  year = 2025,
  journal = {Diversitas Journal},
  volume = {10},
  number = {special\_1},
  doi = {10.48017/dj.v10ispecial_1.3151},
  url = {https://diversitasjournal.com.br/diversitas_journal/article/view/3151}
}

@article{lucapajolaNovelReviewHelpfulness2023,
  title = {A Novel Review Helpfulness Measure Based on the User-Review-Item Paradigm},
  author = {{Luca Pajola} and {Dongkai Chen} and {Mauro Conti} and {V.S. Subrahmanian}},
  year = 2023,
  journal = {ACM Transactions on the Web},
  volume = {17},
  number = {4},
  pages = {1--31},
  doi = {10.1145/3585280},
  url = {https://dl.acm.org/doi/10.1145/3585280}
}

@article{lucapetruzzellisMobilePhoneChoice2010,
  title = {Mobile Phone Choice: {{Technology}} versus Marketing. {{The}} Brand Effect in the Italian Market},
  shorttitle = {Mobile Phone Choice},
  author = {{Luca Petruzzellis}},
  editor = {{T.C. Melewar}},
  year = 2010,
  journal = {European Journal of Marketing},
  volume = {44},
  number = {5},
  pages = {610--634},
  doi = {10.1108/03090561011032298},
  url = {http://www.emerald.com/ejm/article/44/5/610-634/89404}
}

@article{malaquiasSmartphoneUsersSatisfaction2020,
  title = {Smartphone Users' Satisfaction and Regional Aspects:{{Factors}} That Emerge from Online Reviews},
  shorttitle = {Smartphone {{Users}}\&\#8217; {{Satisfaction}} and {{Regional Aspects}}},
  author = {Malaquias, Fernanda Francielle De Oliveira and Silva J{\'u}nior, Romes Jorge Da},
  year = 2020,
  journal = {Journal of technology management \& innovation},
  volume = {15},
  number = {1},
  pages = {3--14},
  doi = {10.4067/S0718-27242020000100003},
  url = {http://www.scielo.cl/scielo.php?script=sci_arttext&pid=S0718-27242020000100003&lng=en&nrm=iso&tlng=en}
}

@article{maramassiLLMcureLLMbasedCompetitor2025,
  title = {{{LLM-cure}}: {{LLM-based}} Competitor User Review Analysis for Feature Enhancement},
  shorttitle = {{{LLM-Cure}}},
  author = {{Maram Assi} and {Safwat Hassan} and {Ying Zou}},
  year = 2025,
  journal = {ACM Transactions on Software Engineering and Methodology},
  publisher = {Association for Computing Machinery (ACM)},
  doi = {10.1145/3744644},
  url = {https://dl.acm.org/doi/10.1145/3744644}
}

@book{marchassenzahlExperienceDesignTechnology2010,
  title = {Experience Design: {{Technology}} for All the Right Reasons},
  shorttitle = {Experience Design},
  editor = {{Marc Hassenzahl}},
  year = 2010,
  series = {Synthesis Lectures on Human-Centered Informatics},
  publisher = {Morgan \& Claypool Publishers},
  address = {San Rafael},
  isbn = {978-1-60845-047-3 978-1-60845-048-0}
}

@article{marchassenzahlNeedsAffectInteractive2010,
  title = {Needs, Affect, and Interactive Products -- Facets of User Experience},
  author = {{Marc Hassenzahl} and {Sarah Diefenbach} and {Anja G\"oritz}},
  year = 2010,
  journal = {Interacting with Computers},
  volume = {22},
  number = {5},
  pages = {353--362},
  doi = {10.1016/j.intcom.2010.04.002},
  url = {https://academic.oup.com/iwc/article-lookup/doi/10.1016/j.intcom.2010.04.002}
}

@incollection{marchassenzahlThingUnderstandingRelationship2003,
  title = {The Thing and {{I}}: {{Understanding}} the Relationship between User and Product},
  shorttitle = {The {{Thing}} and {{I}}},
  booktitle = {Funology},
  author = {{Marc Hassenzahl}},
  editor = {{John Karat} and {Jean Vanderdonckt} and {Gregory Abowd} and {Ga\"alle Calvary} and {John Carroll} and {Gilbert Cockton} and {Mary Czerwinski} and {Steve Feiner} and {Elizabeth Furtado} and {Kristiana H\"o\"ok} and {Robert Jacob} and {Robin Jeffries} and {Peter Johnson} and {Kumiyo Nakakoji} and {Philippe Palanque} and {Oscar Pastor} and {Fabio Patern\`o} and {Costin Pribeanu} and {Marilyn Salzman} and {Chris Schmandt} and {Markus Stolze} and {Gerd Szwillus} and {Manfred Tscheligi} and {Gerrit Van Der Veer} and {Shumin Zhai} and {Mark A. Blythe} and {Kees Overbeeke} and {Andrew F. Monk} and {Peter C. Wright}},
  year = 2003,
  volume = {3},
  pages = {31--42},
  publisher = {Springer Netherlands},
  address = {Dordrecht},
  doi = {10.1007/1-4020-2967-5_4},
  url = {http://link.springer.com/10.1007/1-4020-2967-5_4},
  isbn = {978-1-4020-2966-0 978-1-4020-2967-7}
}

@inproceedings{mariapontikiSemEval2016Task52016,
  title = {{{SemEval-2016}} Task 5: {{Aspect}} Based Sentiment Analysis},
  shorttitle = {{{SemEval-2016 Task}} 5},
  booktitle = {Proceedings of the 10th {{International Workshop}} on {{Semantic Evaluation}} ({{SemEval-2016}})},
  author = {{Maria Pontiki} and {Dimitris Galanis} and {Haris Papageorgiou} and {Ion Androutsopoulos} and {Suresh Manandhar} and {Mohammad AL-Smadi} and {Mahmoud Al-Ayyoub} and {Yanyan Zhao} and {Bing Qin} and {Orphee De Clercq} and {Veronique Hoste} and {Marianna Apidianaki} and {Xavier Tannier} and {Natalia Loukachevitch} and {Evgeniy Kotelnikov} and {N\'uria Bel} and {Salud Mar\'ia Jim\'enez-Zafra} and {G\"ul\c sen Eryi\u git}},
  year = 2016,
  pages = {19--30},
  publisher = {Association for Computational Linguistics},
  address = {San Diego, California},
  doi = {10.18653/v1/S16-1002},
  url = {http://aclweb.org/anthology/S16-1002}
}

@article{markdereuverDomesticationSmartphonesMobile2016,
  title = {Domestication of Smartphones and Mobile Applications: {{A}} Quantitative Mixed-Method Study},
  shorttitle = {Domestication of Smartphones and Mobile Applications},
  author = {{Mark De Reuver} and {Shahrokh Nikou} and {Harry Bouwman}},
  year = 2016,
  journal = {Mobile Media \& Communication},
  volume = {4},
  number = {3},
  pages = {347--370},
  doi = {10.1177/2050157916649989},
  url = {https://journals.sagepub.com/doi/10.1177/2050157916649989}
}

@article{markusgahlerCustomerExperienceConceptualization2023,
  title = {Customer {{Experience}}: {{Conceptualization}}, {{Measurement}}, and {{Application}} in {{Omnichannel Environments}}},
  shorttitle = {Customer {{Experience}}},
  author = {{Markus Gahler} and {Jan F. Klein} and {Michael Paul}},
  year = 2023,
  journal = {Journal of Service Research},
  volume = {26},
  number = {2},
  pages = {191--211},
  doi = {10.1177/10946705221126590},
  url = {https://journals.sagepub.com/doi/10.1177/10946705221126590}
}

@article{maxhortSurveyPerformanceOptimization2022,
  title = {A Survey of Performance Optimization for Mobile Applications},
  author = {{Max Hort} and {Maria Kechagia} and {Federica Sarro} and {Mark Harman}},
  year = 2022,
  journal = {IEEE Transactions on Software Engineering},
  volume = {48},
  number = {8},
  pages = {2879--2904},
  doi = {10.1109/TSE.2021.3071193},
  url = {https://ieeexplore.ieee.org/document/9397392/}
}

@article{mohammedaldeenEndusersKnowBest2024,
  title = {End-Users Know Best: {{Identifying}} Undesired Behavior of Alexa Skills through User Review Analysis},
  shorttitle = {End-{{Users Know Best}}},
  author = {{Mohammed Aldeen} and {Jeffrey Young} and {Song Liao} and {Tsu-Yao Chang} and {Long Cheng} and {Haipeng Cai} and {Xiapu Luo} and {Hongxin Hu}},
  year = 2024,
  journal = {Proceedings of the ACM on Interactive, Mobile, Wearable and Ubiquitous Technologies},
  volume = {8},
  number = {3},
  pages = {1--28},
  doi = {10.1145/3678517},
  url = {https://dl.acm.org/doi/10.1145/3678517}
}

@article{nanfengEffectsReviewSpam2018,
  title = {Effects of Review Spam in a Firm-Initiated Virtual Brand Community: {{Evidence}} from Smartphone Customers},
  shorttitle = {Effects of Review Spam in a Firm-Initiated Virtual Brand Community},
  author = {{Nan Feng} and {Zhenjing Su} and {Dahui Li} and {Chundong Zheng} and {Minqiang Li}},
  year = 2018,
  journal = {Information \& Management},
  volume = {55},
  number = {8},
  pages = {1061--1070},
  doi = {10.1016/j.im.2018.05.012},
  url = {https://linkinghub.elsevier.com/retrieve/pii/S037872061730232X}
}

@inproceedings{nilskluwerContextCategoriesImplementing2025,
  title = {Context over Categories: {{Implementing}} the Theory of Constructed Emotion with {{LLM-guided}} User Analysis},
  shorttitle = {Context over {{Categories}}},
  booktitle = {Proceedings of the {{Extended Abstracts}} of the {{CHI Conference}} on {{Human Factors}} in {{Computing Systems}}},
  author = {{Nils Kl\"uwer} and {Irina Nalis} and {Julia Neidhardt}},
  year = 2025,
  pages = {1--7},
  publisher = {ACM},
  address = {Yokohama Japan},
  doi = {10.1145/3706599.3721205},
  url = {https://dl.acm.org/doi/10.1145/3706599.3721205},
  isbn = {979-8-4007-1395-8}
}

@inproceedings{nimashaarambepolaFactorsInfluencingMobile2024,
  title = {Factors Influencing Mobile App User Experience: {{An}} Analysis of Education App User Reviews},
  shorttitle = {Factors {{Influencing Mobile App User Experience}}},
  booktitle = {2024 4th {{International Conference}} on {{Advanced Research}} in {{Computing}} ({{ICARC}})},
  author = {{Nimasha Arambepola} and {Lankeshwara Munasinghe} and {Nalin Warnajith}},
  year = 2024,
  pages = {223--228},
  publisher = {IEEE},
  address = {Belihuloya, Sri Lanka},
  doi = {10.1109/icarc61713.2024.10499727},
  url = {https://ieeexplore.ieee.org/document/10499727/}
}

@article{odednetzerMineYourOwn2012,
  title = {Mine Your Own Business: {{Market-structure}} Surveillance through Text Mining},
  shorttitle = {Mine {{Your Own Business}}},
  author = {{Oded Netzer} and {Ronen Feldman} and {Jacob Goldenberg} and {Moshe Fresko}},
  year = 2012,
  journal = {Marketing Science},
  volume = {31},
  number = {3},
  pages = {521--543},
  publisher = {{Institute for Operations Research and the Management Sciences (INFORMS)}},
  doi = {10.1287/mksc.1120.0713},
  url = {https://pubsonline.informs.org/doi/10.1287/mksc.1120.0713}
}

@techreport{paddlenlpcontributorsPaddleNLPEasytouseHigh2021,
  title = {{{PaddleNLP}}: {{An}} Easy-to-Use and High Performance {{NLP}} Library},
  author = {{PaddleNLP Contributors}},
  year = 2021,
  url = {https://github.com/PaddlePaddle/PaddleNLP}
}

@misc{pawelweichbrothUsabilityIssuesMobile2025,
  title = {Usability Issues with Mobile Applications: {{Insights}} from Practitioners and Future Research Directions},
  shorttitle = {Usability {{Issues With Mobile Applications}}},
  author = {{Pawel Weichbroth}},
  year = 2025,
  publisher = {arXiv},
  doi = {10.48550/ARXIV.2502.05120},
  url = {https://arxiv.org/abs/2502.05120}
}

@book{philipkotlerMarketingManagement2016,
  title = {Marketing Management},
  author = {{Philip Kotler} and {Kevin Lane Keller}},
  year = 2016,
  edition = {15 [edition]},
  publisher = {Pearson},
  address = {Boston},
  isbn = {978-0-13-385646-0},
  lccn = {HF5415.13 .K64 2016}
}

@inproceedings{qingxiaozhengEvAlignUXAdvancingUX2025,
  title = {{{EvAlignUX}}: {{Advancing UX}} Evaluation through {{LLM-supported}} Metrics Exploration},
  shorttitle = {{{EvAlignUX}}},
  booktitle = {Proceedings of the 2025 {{CHI Conference}} on {{Human Factors}} in {{Computing Systems}}},
  author = {{Qingxiao Zheng} and {Minrui Chen} and {Pranav Sharma} and {Yiliu Tang} and {Mehul Oswal} and {Yiren Liu} and {Yun Huang}},
  year = 2025,
  pages = {1--25},
  publisher = {ACM},
  address = {Yokohama Japan},
  doi = {10.1145/3706598.3714045},
  url = {https://dl.acm.org/doi/10.1145/3706598.3714045},
  isbn = {979-8-4007-1394-1}
}

@inproceedings{rahularalikatteFaultYourStars2018,
  title = {Fault in Your Stars: {{An}} Analysis of Android App Reviews},
  shorttitle = {Fault in Your Stars},
  booktitle = {Proceedings of the {{ACM India Joint International Conference}} on {{Data Science}} and {{Management}} of {{Data}}},
  author = {{Rahul Aralikatte} and {Giriprasad Sridhara} and {Neelamadhav Gantayat} and {Senthil Mani}},
  year = 2018,
  pages = {57--66},
  publisher = {ACM},
  address = {Goa India},
  doi = {10.1145/3152494.3152500},
  url = {https://dl.acm.org/doi/10.1145/3152494.3152500}
}

@article{renevieirasantinSystematicReviewAspectbased2025,
  title = {Systematic Review on Aspect-Based Sentiment Analysis in Cross-Domain},
  author = {{Ren\'e Vieira Santin} and {Solange Oliveira Rezende}},
  year = 2025,
  journal = {Artificial Intelligence Review},
  volume = {59},
  number = {1},
  pages = {37},
  doi = {10.1007/s10462-025-11437-x},
  url = {https://link.springer.com/10.1007/s10462-025-11437-x}
}

@article{russellw.belkExtendedSelfDigital2013,
  title = {Extended Self in a Digital World},
  shorttitle = {Extended {{Self}} in a {{Digital World}}},
  author = {{Russell W. Belk}},
  year = 2013,
  journal = {Journal of Consumer Research},
  volume = {40},
  number = {3},
  pages = {477--500},
  doi = {10.1086/671052},
  url = {https://academic.oup.com/jcr/article-lookup/doi/10.1086/671052}
}

@article{russellw.belkPossessionsExtendedSelf1988,
  title = {Possessions and the Extended Self},
  author = {{Russell W. Belk}},
  year = 1988,
  journal = {Journal of Consumer Research},
  volume = {15},
  number = {2},
  pages = {139--168},
  doi = {10.1086/209154},
  url = {https://academic.oup.com/jcr/article/15/2/139/1841428}
}

@misc{saibogengJSONSchemaBenchRigorousBenchmark2025,
  title = {{{JSONSchemaBench}}: {{A}} Rigorous Benchmark of Structured Outputs for Language Models},
  shorttitle = {{{JSONSchemaBench}}},
  author = {{Saibo Geng} and {Hudson Cooper} and {Micha\l{} Moskal} and {Samuel Jenkins} and {Julian Berman} and {Nathan Ranchin} and {Robert West} and {Eric Horvitz} and {Harsha Nori}},
  year = 2025,
  publisher = {arXiv},
  doi = {10.48550/ARXIV.2501.10868},
  url = {https://arxiv.org/abs/2501.10868}
}

@article{salemalghamdiAspectbasedSentimentAnalysis2024,
  title = {Aspect-Based Sentiment Analysis in Smart Devices: {{A}} Comprehensive and Specialized Dataset},
  shorttitle = {Aspect-Based Sentiment Analysis in Smart Devices},
  author = {{Salem Alghamdi} and {Yaser Alhasawi}},
  year = 2024,
  journal = {Data in Brief},
  volume = {55},
  pages = {110642},
  doi = {10.1016/j.dib.2024.110642},
  url = {https://linkinghub.elsevier.com/retrieve/pii/S2352340924006097}
}

@inproceedings{sarahdiefenbachSmartphonePacifierIts2019,
  title = {The Smartphone as a Pacifier and Its Consequences: {{Young}} Adults' Smartphone Usage in Moments of Solitude and Correlations to Self-Reflection},
  shorttitle = {The {{Smartphone}} as a {{Pacifier}} and Its {{Consequences}}},
  booktitle = {Proceedings of the 2019 {{CHI Conference}} on {{Human Factors}} in {{Computing Systems}}},
  author = {{Sarah Diefenbach} and {Kim Borrmann}},
  year = 2019,
  pages = {1--14},
  publisher = {ACM},
  address = {Glasgow Scotland Uk},
  doi = {10.1145/3290605.3300536},
  url = {https://dl.acm.org/doi/10.1145/3290605.3300536},
  isbn = {978-1-4503-5970-2}
}

@article{satanikmitraOBIMComputationalModel2020,
  title = {{{OBIM}}: {{A}} Computational Model to Estimate Brand Image from Online Consumer Review},
  shorttitle = {{{OBIM}}},
  author = {{Satanik Mitra} and {Mamata Jenamani}},
  year = 2020,
  journal = {Journal of Business Research},
  volume = {114},
  pages = {213--226},
  doi = {10.1016/j.jbusres.2020.04.003},
  url = {https://linkinghub.elsevier.com/retrieve/pii/S0148296320302204}
}

@article{sebastiana.c.perrigSmartphoneAppAesthetics2023,
  title = {Smartphone App Aesthetics Influence Users' Experience and Performance},
  author = {{Sebastian A. C. Perrig} and {David Ueffing} and {Klaus Opwis} and {Florian Br\"uhlmann}},
  year = 2023,
  journal = {Frontiers in Psychology},
  volume = {14},
  pages = {1113842},
  doi = {10.3389/fpsyg.2023.1113842},
  url = {https://www.frontiersin.org/articles/10.3389/fpsyg.2023.1113842/full}
}

@inproceedings{soussandjamasbiMUXDevelopmentHolistic2017,
  title = {{{MUX}}: {{Development}} of a Holistic Mobile User Experience Instrument},
  shorttitle = {{{MUX}}},
  booktitle = {Hawaii {{International Conference}} on {{System Sciences}}},
  author = {{Soussan Djamasbi} and {Vance Wilson}},
  year = 2017,
  eprint = {10125/41217},
  eprinttype = {hdl},
  doi = {10.24251/HICSS.2017.068},
  url = {http://hdl.handle.net/10125/41217}
}

@article{tunthurathetAspectbasedSentimentAnalysis2010,
  title = {Aspect-Based Sentiment Analysis of Movie Reviews on Discussion Boards},
  author = {{Tun Thura Thet} and {Jin-Cheon Na} and {Christopher S.G. Khoo}},
  year = 2010,
  journal = {Journal of Information Science},
  volume = {36},
  number = {6},
  pages = {823--848},
  doi = {10.1177/0165551510388123},
  url = {https://journals.sagepub.com/doi/10.1177/0165551510388123}
}

@article{waltert.nakamuraWhatFactorsAffect2022,
  title = {What Factors Affect the {{UX}} in Mobile Apps? {{A}} Systematic Mapping Study on the Analysis of App Store Reviews},
  shorttitle = {What Factors Affect the {{UX}} in Mobile Apps?},
  author = {{Walter T. Nakamura} and {Edson Cesar De Oliveira} and {Elaine H.T. De Oliveira} and {David Redmiles} and {Tayana Conte}},
  year = 2022,
  journal = {Journal of Systems and Software},
  volume = {193},
  pages = {111462},
  doi = {10.1016/j.jss.2022.111462},
  url = {https://linkinghub.elsevier.com/retrieve/pii/S0164121222001509}
}

@article{weifenghuResearchBrandImage2023,
  title = {Research on the Brand Image of {{iOS}} and Android Smart Phone Operating Systems Based on Mixed Methods},
  author = {{Weifeng Hu} and {Tianyun Hao} and {Yue Hu} and {Hui Chen} and {Yi Zhou} and {Wantong Yin}},
  year = 2023,
  journal = {Frontiers in Psychology},
  volume = {14},
  pages = {1040180},
  doi = {10.3389/fpsyg.2023.1040180},
  url = {https://www.frontiersin.org/articles/10.3389/fpsyg.2023.1040180/full}
}

@inproceedings{wenxuanzhangSentimentAnalysisEra2024,
  title = {Sentiment Analysis in the Era of Large Language Models: {{A}} Reality Check},
  shorttitle = {Sentiment {{Analysis}} in the {{Era}} of {{Large Language Models}}},
  booktitle = {Findings of the {{Association}} for {{Computational Linguistics}}: {{NAACL}} 2024},
  author = {{Wenxuan Zhang} and {Yue Deng} and {Bing Liu} and {Sinno Pan} and {Lidong Bing}},
  year = 2024,
  pages = {3881--3906},
  publisher = {Association for Computational Linguistics},
  address = {Mexico City, Mexico},
  doi = {10.18653/v1/2024.findings-naacl.246},
  url = {https://aclanthology.org/2024.findings-naacl.246}
}

@article{wenxuanzhangSurveyAspectbasedSentiment2023,
  title = {A Survey on Aspect-Based Sentiment Analysis: {{Tasks}}, Methods, and Challenges},
  shorttitle = {A {{Survey}} on {{Aspect-Based Sentiment Analysis}}},
  author = {{Wenxuan Zhang} and {Xin Li} and {Yang Deng} and {Lidong Bing} and {Wai Lam}},
  year = 2023,
  journal = {IEEE Transactions on Knowledge and Data Engineering},
  volume = {35},
  number = {11},
  pages = {11019--11038},
  doi = {10.1109/TKDE.2022.3230975},
  url = {https://ieeexplore.ieee.org/document/9996141/}
}

@article{whitney-jocelynkouahoInvestigatingPerspectivesExperiences2024,
  title = {Investigating Perspectives of and Experiences with Low Cost Commercial Fitness Wearables},
  author = {{Whitney-Jocelyn Kouaho} and {Daniel A. Epstein}},
  year = 2024,
  journal = {Proceedings of the ACM on Interactive, Mobile, Wearable and Ubiquitous Technologies},
  volume = {8},
  number = {4},
  pages = {1--22},
  doi = {10.1145/3699740},
  url = {https://dl.acm.org/doi/10.1145/3699740}
}

@inproceedings{xinshengxuApproachExtractProduct2017,
  title = {An Approach to Extract Product Features from {{Chinese}} Consumer Reviews and Establish Product Feature Structure Tree},
  booktitle = {International {{Journal}} of {{Computational Linguistics}} \& {{Chinese Language Processing}}, {{Volume}} 22, {{Number}} 1, {{June}} 2017},
  author = {{Xinsheng Xu} and {Jing Lin} and {Ying Xiao} and {Jianzhe Yu}},
  editor = {{Yuen-Hsien Tseng} and {Jen-Tzung Chien}},
  year = 2017,
  url = {https://aclanthology.org/O17-2003/}
}

@inproceedings{yeyiranAspectbasedSentimentAnalysis2019,
  title = {Aspect-Based Sentiment Analysis on Mobile Phone Reviews with {{LDA}}},
  booktitle = {Proceedings of the 2019 4th {{International Conference}} on {{Machine Learning Technologies}}},
  author = {{Ye Yiran} and {Sangeet Srivastava}},
  year = 2019,
  pages = {101--105},
  publisher = {ACM},
  address = {Nanchang China},
  doi = {10.1145/3340997.3341012},
  url = {https://dl.acm.org/doi/10.1145/3340997.3341012},
  isbn = {978-1-4503-6323-5}
}

@article{youngryuDecisionModelsComparative2007,
  title = {Decision Models for Comparative Usability Evaluation of Mobile Phones Using the Mobile Phone Usability Questionnaire ({{MPUQ}})},
  author = {{Young Ryu} and {Kari Babski-Reeves} and {Tonya Smith-Jackson} and {Maury A. Nussbaum}},
  year = 2007,
  journal = {Journal of Usability Studies},
  volume = {3},
  number = {1},
  pages = {24--39},
  publisher = {Usability Professionals' Association},
  address = {Bloomingdale, IL}
}

@article{youngsamryuReliabilityValidityMobile2006,
  title = {Reliability and Validity of the Mobile Phone Usability Questionnaire ({{MPUQ}})},
  author = {{Young Sam Ryu} and {Tonya L. Smith-Jackson}},
  year = 2006,
  journal = {Journal of Usability Studies},
  volume = {2},
  number = {1},
  pages = {39--53},
  publisher = {Usability Professionals' Association},
  address = {Bloomingdale, IL},
  url = {https://dl.acm.org/doi/abs/10.5555/2835536.2835540}
}

@article{yuanchunliMiningUserReviews2017,
  title = {Mining User Reviews for Mobile App Comparisons},
  author = {{Yuanchun Li} and {Baoxiong Jia} and {Yao Guo} and {Xiangqun Chen}},
  year = 2017,
  journal = {Proceedings of the ACM on Interactive, Mobile, Wearable and Ubiquitous Technologies},
  volume = {1},
  number = {3},
  pages = {1--15},
  doi = {10.1145/3130935},
  url = {https://dl.acm.org/doi/10.1145/3130935}
}

@article{yuwangProductCompetitiveAnalysis2025,
  title = {Product Competitive Analysis Model Based on Consumer Preference Satisfaction Similarity: {{Case}} Study of Smartphone {{UGC}}},
  shorttitle = {Product {{Competitive Analysis Model Based}} on {{Consumer Preference Satisfaction Similarity}}},
  author = {{Yu Wang} and {Jiacong Wu} and {Xu Ye} and {Yue Wu}},
  year = 2025,
  journal = {Systems},
  volume = {13},
  number = {1},
  pages = {38},
  doi = {10.3390/systems13010038},
  url = {https://www.mdpi.com/2079-8954/13/1/38}
}

@misc{zengzhiwangChatGPTGoodSentiment2023,
  title = {Is {{ChatGPT}} a Good Sentiment Analyzer? {{A}} Preliminary Study},
  shorttitle = {Is {{ChatGPT}} a {{Good Sentiment Analyzer}}?},
  author = {{Zengzhi Wang} and {Qiming Xie} and {Yi Feng} and {Zixiang Ding} and {Zinong Yang} and {Rui Xia}},
  year = 2023,
  publisher = {arXiv},
  doi = {10.48550/ARXIV.2304.04339},
  url = {https://arxiv.org/abs/2304.04339}
}

@article{danielhintzeLargescaleLongtermAnalysis2017,
  title = {A Large-Scale, Long-Term Analysis of Mobile Device Usage Characteristics},
  author = {{Daniel Hintze} and {Philipp Hintze} and {Rainhard D. Findling} and {René Mayrhofer}},
  date = {2017-06-30},
  journaltitle = {Proceedings of the ACM on Interactive, Mobile, Wearable and Ubiquitous Technologies},
  shortjournal = {Proc. ACM Interact. Mob. Wearable Ubiquitous Technol.},
  volume = {1},
  number = {2},
  pages = {1--21},
  doi = {10.1145/3090078},
  url = {https://dl.acm.org/doi/10.1145/3090078}
}

@incollection{rusuCustomerEXperienceBridge2020a,
  title = {Customer {{eXperience}}: {{A}} Bridge between Service Science and Human-Computer Interaction},
  shorttitle = {Customer {{eXperience}}},
  booktitle = {Human {{Systems Engineering}} and {{Design II}}},
  author = {{Virginica Rusu} and {Cristian Rusu} and {Federico Botella} and {Daniela Quiñones} and {Camila Bascur} and {Virginia Zaraza Rusu}},
  editor = {{Tareq Ahram} and {Waldemar Karwowski} and {Stefan Pickl} and {Redha Taiar}},
  date = {2020},
  volume = {1026},
  pages = {385--390},
  publisher = {Springer International Publishing},
  location = {Cham},
  doi = {10.1007/978-3-030-27928-8_59},
  url = {http://link.springer.com/10.1007/978-3-030-27928-8_59},
  isbn = {978-3-030-27927-1 978-3-030-27928-8}
}

@article{quinonesUnderstandingCustomerExperience2023,
  title = {Understanding the Customer Experience in Human-Computer Interaction: {{A}} Systematic Literature Review},
  shorttitle = {Understanding the Customer Experience in Human-Computer Interaction},
  author = {{Daniela Quiñones} and {Luis Rojas}},
  date = {2023-02-10},
  journaltitle = {PeerJ Computer Science},
  volume = {9},
  pages = {e1219},
  doi = {10.7717/peerj-cs.1219},
  url = {https://peerj.com/articles/cs-1219}
}

@article{katherinen.lemonUnderstandingCustomerExperience2016,
  title = {Understanding Customer Experience throughout the Customer Journey},
  author = {{Katherine N. Lemon} and {Peter C. Verhoef}},
  date = {2016-11},
  journaltitle = {Journal of Marketing},
  shortjournal = {Journal of Marketing},
  volume = {80},
  number = {6},
  pages = {69--96},
  doi = {10.1509/jm.15.0420},
  url = {https://journals.sagepub.com/doi/10.1509/jm.15.0420}
}

@article{gentileHowSustainCustomer2007,
  title = {How to Sustain the Customer Experience: {{An}} Overview of Experience Components That Co-Create Value with the Customer},
  shorttitle = {How to {{Sustain}} the {{Customer Experience}}},
  author = {{Chiara Gentile} and {Nicola Spiller} and {Giuliano Noci}},
  date = {2007-10},
  journaltitle = {European Management Journal},
  shortjournal = {European Management Journal},
  volume = {25},
  number = {5},
  pages = {395--410},
  doi = {10.1016/j.emj.2007.08.005},
  url = {https://linkinghub.elsevier.com/retrieve/pii/S0263237307000886}
}

@article{haynesContentValidityPsychological1995,
  title = {Content Validity in Psychological Assessment: {{A}} Functional Approach to Concepts and Methods.},
  shorttitle = {Content Validity in Psychological Assessment},
  author = {{Stephen N. Haynes} and {David C. S. Richard} and {Edward S. Kubany}},
  date = {1995-09},
  journaltitle = {Psychological Assessment},
  shortjournal = {Psychological Assessment},
  volume = {7},
  number = {3},
  pages = {238--247},
  doi = {10.1037/1040-3590.7.3.238},
  url = {https://doi.apa.org/doi/10.1037/1040-3590.7.3.238}
}

@article{hsuDelphiTechniqueMaking,
  title = {The Delphi Technique: {{Making}} Sense of Consensus},
  shorttitle = {The {{Delphi Technique}}},
  author = {{Chia-Chien Hsu} and {Brian A. Sandford}},
  date = {2007},
  publisher = {University of Massachusetts Amherst},
  doi = {10.7275/PDZ9-TH90},
  url = {https://openpublishing.library.umass.edu/pare/article/id/1418/}
}

@book{linstoneDelphiMethodTechniques1975,
  title = {The Delphi Method: {{Techniques}} and Applications},
  shorttitle = {The {{Delphi}} Method},
  editor = {{Harold A. Linstone} and {Murray Turoff}},
  date = {1975},
  publisher = {Addison-Wesley},
  location = {Reading, Mass.},
  isbn = {978-0-201-04294-8 978-0-201-04293-1},
  pagetotal = {620}
}


\section*{Appendix}
\addcontentsline{toc}{chapter}{Appendix}


\appendix

\section{Review Panel Details}
\label{Experts Information}
\revise{
To protect participant anonymity while providing additional transparency about the panel-based review, we summarize the panel composition, relevant background, methodological or project experience, and review roles in Table~\ref{tab:expertPanel}. The table reports grouped information rather than personally identifiable details. This grouped presentation shows how different review roles contributed complementary theoretical, methodological, domain-informed, and product-facing perspectives to the refinement of the framework.
}

\begin{table}[ht]
    \caption{\revise{Composition and relevant experience of the review panel.}}
    \label{tab:expertPanel}
    \centering
    \small
    \renewcommand{\arraystretch}{1.2}
    \setlength{\tabcolsep}{4pt}

    \begin{tabularx}{\columnwidth}{
        >{\RaggedRight\arraybackslash}p{0.17\columnwidth}
        >{\RaggedRight\arraybackslash}p{0.04\columnwidth}
        >{\RaggedRight\arraybackslash}p{0.25\columnwidth}
        >{\RaggedRight\arraybackslash}X
    }

    \toprule

    \revise{\textbf{Review role}}
    & \revise{\textbf{N}}
    & \revise{\textbf{Relevant background}}
    & \revise{\textbf{Relevant expertise and review contribution}} \\

    \midrule

    \revise{Conceptual framework reviewers}
    & \revise{4}
    & \revise{UX evaluation, consumer behavior, psychology/user evaluation, and sociology of technology use}
    & \revise{Experience evaluation, consumer experience, construct interpretation, and measurement-related research involving experience assessment and user evaluation. Reviewed conceptual coverage, metric definitions, hierarchical coherence, and construct boundaries.} \\

    \addlinespace

    \revise{Research and coding reviewers}
    & \revise{2}
    & \revise{UX research, product experience, consumer experience, and technology use research}
    & \revise{User studies, qualitative coding, taxonomy refinement, survey-based user evaluation, and experience assessment. Reviewed metric boundaries, coding clarity, and consistency between definitions and examples.} \\

    \addlinespace

    \revise{Smartphone review and product experience reviewers}
    & \revise{4}
    & \revise{Smartphone product experience, user-generated content analysis, interaction experience, and social experience}
    & \revise{Smartphone product analysis, review interpretation, user-generated content coding, and experience-oriented data analysis. Assessed whether metric definitions were understandable and applicable to smartphone review evidence.} \\

    \addlinespace

    \revise{Smartphone industry practitioners}
    & \revise{3}
    & \revise{Smartphone product UX, mobile product evaluation, user research, and product experience}
    & \revise{Professional experience in smartphone UX analysis, product diagnosis, product requirement analysis, and metric interpretation. Reviewed product-facing interpretability, practical relevance, terminology, and actionability.} \\

    \bottomrule

    \end{tabularx}
\end{table}

\section{The Hierarchical Smartphone UX Measurement Framework (L1-L4)}
\label{4-level metrics}
\subsection{Functional Experience Metrics and Definitions}

\begin{table}[ht]
\caption{All sub-metrics of the Functional Experience metric.}
\label{tab:FE}
\resizebox{0.9\textwidth}{!}{
\begin{tabular}{|l|c|c|c|}
\hline
\rowcolor[HTML]{C0C0C0} 
Level-1 metrics & Level-2 metrics & Level-3 metrics & Level-4 metrics \\ \hline
\multirow{42}{*}{Functional Experience (FE)} & \multirow{12}{*}{Usefulness}  & \multirow{2}{*}{Time Saving}                & Accuracy of Personalized Recommendations                \\ \cline{4-4} 
                                             &                                    &                                                & Application Compatibility                               \\ \cline{3-4} 
                                             &                                    & \multirow{4}{*}{Labor Saving}            & Cross-Device Collaboration                              \\ \cline{4-4} 
                                             &                                    &                                                & Multi-Scenario Applicability                            \\ \cline{4-4} 
                                             &                                    &                                                & Adaptability to Physical Conditions                     \\ \cline{4-4} 
                                             &                                    &                                                & Quick Operations                                        \\ \cline{3-4} 
                                             &                                    & \multirow{6}{*}{Increase Usage Benefits} & Smoothness                                              \\ \cline{4-4} 
                                             &                                    &                                                & Intelligence                                            \\ \cline{4-4} 
                                             &                                    &                                                & Functional   Innovativeness                             \\ \cline{4-4} 
                                             &                                    &                                                & Functional Autonomy                                     \\ \cline{4-4} 
                                             &                                    &                                                & Functional   Playfulness                                \\ \cline{4-4} 
                                             &                                    &                                                & Functional Richness                                     \\ \cline{2-4} 
                                             & \multirow{8}{*}{Ease of Use}    & \multirow{2}{*}{Learnability}             & Learning Threshold                                      \\ \cline{4-4} 
                                             &                                    &                                                & Learning Cost                                           \\ \cline{3-4} 
                                             &                                    & \multirow{2}{*}{Clarity}                  & Logical Consistency                                     \\ \cline{4-4} 
                                             &                                    &                                                & Layout Rationality                                      \\ \cline{3-4} 
                                             &                                    & \multirow{4}{*}{Operability}              & Task Execution   Efficiency                             \\ \cline{4-4} 
                                             &                                    &                                                & Error Tolerance                                         \\ \cline{4-4} 
                                             &                                    &                                                & Memorability                                            \\ \cline{4-4} 
                                             &                                    &                                                & Care for Special Groups (Elderly, Children, Disabled) \\ \cline{2-4} 
                                             & \multirow{11}{*}{Reliability} & \multirow{7}{*}{Stability}                & System Stability                                        \\ \cline{4-4} 
                                             &                                    &                                                & Network Stability                                       \\ \cline{4-4} 
                                             &                                    &                                                & Call Stability                                          \\ \cline{4-4} 
                                             &                                    &                                                & Connection Stability                                    \\ \cline{4-4} 
                                             &                                    &                                                & Charging Stability                                      \\ \cline{4-4} 
                                             &                                    &                                                & Native Software   Stability                             \\ \cline{4-4} 
                                             &                                    &                                                & Third-Party Software   Stability                        \\ \cline{3-4} 
                                             &                                    & \multirow{2}{*}{Security}                & Reduction of Hassles                                    \\ \cline{4-4} 
                                             &                                    &                                                & Risk Reduction                                          \\ \cline{3-4} 
                                             &                                    & \multirow{2}{*}{Responsiveness}           & Service Response   Speed                                \\ \cline{4-4} 
                                             &                                    &                                                & Service Response   Quality                              \\ \cline{2-4} 
                                             & \multirow{11}{*}{Durability}  & \multirow{7}{*}{Battery Life}             & Battery Health   Management                             \\ \cline{4-4} 
                                             &                                    &                                                & Standby Battery Life                                    \\ \cline{4-4} 
                                             &                                    &                                                & Gaming Battery Life                                     \\ \cline{4-4} 
                                             &                                    &                                                & Daily Usage Battery   Life                              \\ \cline{4-4} 
                                             &                                    &                                                & Audio-Video Playback   Battery Life                     \\ \cline{4-4} 
                                             &                                    &                                                & Photography Battery   Life                              \\ \cline{4-4} 
                                             &                                    &                                                & Battery Life in   Power-Saving Mode                     \\ \cline{3-4} 
                                             &                                    & \multirow{2}{*}{Memory Management}        & RAM Management                                          \\ \cline{4-4} 
                                             &                                    &                                                & Storage Management                                      \\ \cline{3-4} 
                                             &                                    & \multirow{2}{*}{Performance Retention}    & Performance   Maintenance                               \\ \cline{4-4} 
                                             &                                    &                                                & System Update                                           \\ \hline
\end{tabular}

}

\end{table}

Functional Experience concerns whether the features provided by a smartphone meet users' needs and whether these features are easy to use and access. It covers the core capabilities of the operating system, such as multitasking, application compatibility, and system stability (Table \ref{tab:FE}). For instance, whether the phone offers a wide range of built-in functions or allows users to conveniently download required applications from an app store are both critical factors in evaluating functional experience.

\subsection{Sensory Experience Metrics and Definitions}
Sensory Experience concerns whether a smartphone's design provides users with intuitive, pleasant, and comfortable perceptions across the visual, auditory, and haptic modalities (Table \ref{tab:SSE}). An operating system that delivers high-quality sensory experience is characterized by appealing interface design, harmonious color schemes, smooth animations, immersive audio, and precise, comfortable touch interactions.
\begin{table}[ht]
\caption{All sub-metrics of the Sensory Experience metric.}
\label{tab:SSE}
\resizebox{0.9\textwidth}{!}{
\begin{tabular}{|l|c|c|c|}
\hline
\rowcolor[HTML]{C0C0C0} 
Level-1 metrics & Level-2 metrics & Level-3 metrics & Level-4 metrics \\ \hline
\multirow{37}{*}{Sensory   Experience (SSE)} & \multirow{18}{*}{Visual   Appeal} & \multirow{8}{*}{Appearance and   Design}   & System Theme   Attractiveness                   \\ \cline{4-4} 
                                             &                                        &                                                 & Font Attractiveness                             \\ \cline{4-4} 
                                             &                                        &                                                 & Wallpaper   Attractiveness                      \\ \cline{4-4} 
                                             &                                        &                                                 & Screen-Off   Attractiveness                     \\ \cline{4-4} 
                                             &                                        &                                                 & Home Screen   Attractiveness                    \\ \cline{4-4} 
                                             &                                        &                                                 & Lock Screen   Attractiveness                    \\ \cline{4-4} 
                                             &                                        &                                                 & Size and Weight                                 \\ \cline{4-4} 
                                             &                                        &                                                 & Appearance                                      \\ \cline{3-4} 
                                             &                                        & \multirow{3}{*}{Display Quality}                & Basic Display   Performance                     \\ \cline{4-4} 
                                             &                                        &                                                 & Brightness Display   Performance                \\ \cline{4-4} 
                                             &                                        &                                                 & Display Performance   under Different Modes     \\ \cline{3-4} 
                                             &                                        & \multirow{5}{*}{Image Aesthetics}               & Camera Performance                              \\ \cline{4-4} 
                                             &                                        &                                                 & Video Stability                                 \\ \cline{4-4} 
                                             &                                        &                                                 & Real-Time Filters and   Beautification Features \\ \cline{4-4} 
                                             &                                        &                                                 & Low-Light Photography                           \\ \cline{4-4} 
                                             &                                        &                                                 & Color Reproduction                              \\ \cline{3-4} 
                                             &                                        & \multirow{2}{*}{Animation Effects}              & Animation Naturalness                           \\ \cline{4-4} 
                                             &                                        &                                                 & Animation Style                                 \\ \cline{2-4} 
                                             & \multirow{9}{*}{Auditory Effect}       & Volume                                          &                                                 \\ \cline{3-4} 
                                             &                                        & Sound Quality                                   &                                                 \\ \cline{3-4} 
                                             &                                        & \multirow{2}{*}{Call Audio   Quality}           & Noise Reduction   Effect                        \\ \cline{4-4} 
                                             &                                        &                                                 & Call Clarity                                    \\ \cline{3-4} 
                                             &                                        & \multirow{3}{*}{Media Audio   Quality}          & Video Audio Quality                             \\ \cline{4-4} 
                                             &                                        &                                                 & Music Audio Quality                             \\ \cline{4-4} 
                                             &                                        &                                                 & Gaming Audio Quality                            \\ \cline{3-4} 
                                             &                                        & \multirow{2}{*}{System Audio   Quality}         & Notification Audio   Quality                    \\ \cline{4-4} 
                                             &                                        &                                                 & Alert Audio Quality                             \\ \cline{2-4} 
                                             & \multirow{10}{*}{Tactile Effect}        & \multirow{3}{*}{Tactile Feedback   Effect}       & Vibration Stability                             \\ \cline{4-4} 
                                             &                                        &                                                 & Vibration Style                                 \\ \cline{4-4} 
                                             &                                        &                                                 & Vibration Intensity                             \\ \cline{3-4} 
                                             &                                        & \multirow{3}{*}{Thermal   Perception}           & Charging Temperature                            \\ \cline{4-4} 
                                             &                                        &                                                 & Daily Usage   Temperature                       \\ \cline{4-4} 
                                             &                                        &                                                 & Gaming Temperature                              \\ \cline{3-4} 
                                             &                                        & \multirow{3}{*}{Touch Response}                 & Touch Stability                                 \\ \cline{4-4} 
                                             &                                        &                                                 & Touch Accuracy                                  \\ \cline{4-4} 
                                             &                                        &                                                 & Touch Performance   under Special Conditions    \\ \cline{3-4} 
                                             &                                        & Surface Texture                                 &                                                 \\ \hline
\end{tabular}
}

\end{table}

\subsection{Social Experience Metrics and Definitions}
Social Experience concerns how the product shapes users’ identity expression and social meaning (Table \ref{tab:SCE}).
The third-level metric Personal Style Expression reflects the extent to which the product enables users to express their personal identity, preferences, and aesthetic style through its design and customizable features. Brand–Self Connectedness reflects the degree to which the product or brand aligns with users’ self-concept, values, and perceived social identity. Self-Improvement reflects how the product supports users in cultivating habits, enhancing skills, or improving their physical or mental well-being during daily use.

Interpersonal Connectedness reflects the extent to which the product facilitates communication, emotional closeness, and meaningful interaction with close others such as family, partners, or friends. Community Participation reflects how the product enables users to join, engage with, or maintain participation in communities that share common interests, norms, or goals. Social Acceptance and Belonging reflects users’ feelings of being recognized, accepted, or socially integrated within broader peer groups or societal contexts through product use.

\begin{table}[ht]
    \caption{All sub-metrics of the Social Experience metric.}
    \label{tab:SCE}
    \centering
    \small
\begin{tabular}{|l|c|c|}
\hline
\rowcolor[HTML]{C0C0C0} 
Level-1 metrics                        & Level-2 metrics                         & Level-3 metrics                     \\ \hline
\multirow{6}{*}{Social Experience} & \multirow{3}{*}{Self-Expression}      & Personal Style Expression       \\ \cline{3-3} 
                                   &                                       & Brand–Self Connectedness        \\ \cline{3-3} 
                                   &                                       & Self-Improvement                \\ \cline{2-3} 
                                   & \multirow{3}{*}{Social Connectedness} & Interpersonal Connectedness     \\ \cline{3-3} 
                                   &                                       & Community Participation         \\ \cline{3-3} 
                                   &                                       & Social Acceptance and Belonging \\ \hline
\end{tabular}

\end{table}

\section{Prompts for Different Tasks}
The following prompts use \{brand\} = Brand A as the target brand. Replace {brand} with the desired brand when applying the prompt.
\subsection{Prompt for Functional and Sensory Experience Extraction Task}
\label{prompt:extrcation}

{\footnotesize
\begin{Prompt}
#Task :Extract localized aspect opinion pairs from a user review about smartphone user experience. The extraction targets Functional Experience and Sensory Experience only. The output must focus on one target brand. Ignore evaluations that apply only to other brands.
##Target brand: The target brand is {brand}.
#Definitions:
##Aspect means a concrete smartphone attribute, system component, interaction element, or perceptual property explicitly mentioned in the text. Examples include battery life, charging, performance, system smoothness, stability, responsiveness, screen, brightness, color, camera, speaker, haptics, temperature, touch response.
##Opinion means the evaluative word or phrase that expresses the user’s judgment about the aspect. It can be an adjective, verb phrase, comparison phrase, or short evaluative clause. Examples include smooth, lags, too hot, not stable, better than, works well, looks great.
##Only extract pairs where the opinion clearly evaluates the aspect in the text. Do not infer missing aspects or opinions.
#Brand filtering rules
Rule 1, extract only evidence units that are about the target brand {brand}.
Rule 2, if the review does not explicitly mention {brand} but clearly refers to the target phone as the main subject, treat the subject as {brand} and extract normally.
Rule 3, if the review mentions {brand} and other brands, extract only the aspect opinion pairs whose subject is {brand}. Ignore aspect opinion pairs that only describe other brands.
Rule 4, if the review only describes other brands and provides no evaluative content about {brand}, output an empty array.
Rule 5, in comparative statements, keep the opinion phrase as it appears, for example better than Brand C, worse than Brand D, but only when the comparison is evaluating {brand}.

#Scope constraints
Include Functional Experience and Sensory Experience content. Exclude social identity relationship meanings, and generic brand praise or dislike without a specific aspect.
#Granularity rules
1. A single review may contain multiple pairs, return all valid pairs about {brand}.
2. If one opinion targets multiple aspects, split into multiple pairs when the mapping is clear.
3. If an aspect is mentioned but there is no evaluation, do not extract it.
4. Keep Aspect as a concise noun phrase. Keep Opinion as the shortest phrase that still preserves the evaluation.
5. Preserve the original language of the review in the output, do not translate.

#Output format
Return only a JSON array of objects. Each object must have exactly three keys: Aspect and Opinion.
##output examples:
Input:
"Brand A is smoother than Brand C, but the Brand C camera is still better."
Output:
[
{"Aspect":"system","Opinion":"smoother than Brand C"}
]

Input:
"The Brand D battery lasts forever. Nothing special about Brand A though."
Output
[
{"Aspect":"overall experience","Opinion":"nothing special"}
]

Input:
OPPO speakers are loud and clear.
Output:
[]
\end{Prompt}
}

\subsection{Prompt for Social Experience Extraction and Classification Task}
\label{prompt:span}
{\footnotesize
\begin{Prompt}
## Role
You are a social psychologist specializing in smartphone consumer behavior, and an expert in qualitative coding.
## Background & Overall Target
Your task is to analyze user reviews about smartphones, identify content related to “Social Experience,” and apply fine-grained labels according to the metric framework provided. The research goal is to determine whether users, during their use of {brand} phones and the operating system, mention anything related to “Social Experience,” and if so, which aspects.
##Target brand: The target brand is {brand}.
Notice: the only brand you should analyze is {brand}, and you only need to capture users’ attitudes toward this brand. If a segment is evaluating another brand, ignore it directly—no further analysis is needed.
“Social Experience” refers to how a smartphone and its operating system affect users’ self-presentation, identity meaning, interpersonal relationships, and sense of group belonging in everyday use. It includes two parts: Self-Expression and Social Connectedness. This task only cares about content related to “Social Experience.” If a passage does not involve any social meaning such as the self, others, groups, identity, belonging, being recognized, etc., it can be labeled as “No relevant dimension.”
## Metric Structure (you must strictly use the following six third-level metric names)
Level 1: Social Experience  
Level 2: Self-Expression: 1)Personal Style Expression; 2)Brand–Self Connectedness; 3)Self-Improvement
Level 2: Social Connectedness: 4) Interpersonal Connectedness; 5) Community Participation; 6) Social Acceptance and Belonging
Below are the definitions and typical expressions for the six third-level metrics. When classifying, you must align strictly with these definitions.
1.  Personal Style Expression:
Reflects the extent to which the product helps users express their personal identity, preferences, and aesthetic style through appearance, UI, photo style, etc., making them feel “this phone really feels like me.”
Typical expressions (examples for understanding only—do NOT copy verbatim):
“Totally matches my style / vibe / aesthetics”
“I can customize themes, icons, and the home screen however I want—very personal”
“It has a certain atmosphere—perfect for me”
Boundary note: comments that only say “looks good” or “beautiful” should NOT be categorized here. It must be “looks good in a way that feels like me / fits me.”
2. Brand–Self Connectedness:
Reflects the extent to which the phone or brand matches the user’s self-concept, values, identity, or social circle, where the user treats the brand as a symbol of their identity or group.
Typical expressions:
“Good for business professionals / students / young people / women”
“This brand feels premium / professional / trustworthy”
“Using Brand D makes me feel national pride,” “Brand A really matches my personality and style”
Boundary note: this is about “whether the brand and I feel like the same kind of people / share values,” not simply “it works well.”
3. Self-Improvement:
Reflects the extent to which the product helps users build habits, improve skills, or enhance physical/mental well-being, making them feel “I’m becoming a better version of myself.”
Typical expressions:
“Study mode / focus mode helps me calm down and study”
“Health check-ins remind me to drink water / exercise / sleep earlier”
“Using schedules and reminders to plan work makes me more disciplined”
Boundary note: this is about the feeling of “self-improvement,” not “better phone performance.”
4. Interpersonal Connectedness:
Reflects the extent to which the product helps users stay connected with close others—family, partners, friends—enhance interaction, and strengthen relationships.
Typical expressions:
“My whole family uses the same brand—communication and photo sharing is so convenient”
“Video calls are so clear—it feels like I’m closer to my parents who live far away”
“Taking photos for my girlfriend / family and recording life makes us feel closer”
Boundary note: only focuses on relationship experiences with specific people (family/partner/friends), not abstract “society.”
5. Community Participation:
Reflects the extent to which the product helps users join, participate in, or sustain a group/community, making them feel “we share a circle / goals / interests.”
Typical expressions:
“In the brand community / user group, people share tips—it has a great vibe”
“I formed a study group with classmates—this phone works great for coordinating”
“Family health cards / study communities help everyone stick to something together”
Boundary note: emphasizes “doing things and participating within a group,” not merely “being recognized by the group.”
6. Social Acceptance and Belonging:
Reflects whether users feel recognized by others, not looked down on, respected, or better able to fit into a group because they use this phone.
Typical expressions:
“It feels classy to take out—doesn’t look cheap or low-end”
“Friends / colleagues complimented my choice”
“This model suits young people / women—using it makes me feel like I fit in”
Boundary note: emphasizes “how others see me” and “my place in the group,” not the function itself.

## Annotation Unit & Multi-Label Rules
First, split the user review into multiple “opinion units,” using: periods, question marks, exclamation marks, semicolons, or clear pivot words (e.g., “but,” “however,” “also,” etc.).
The segmentation does not need to be overly strict; it just needs to keep each unit relatively complete and semantically consistent.
For each opinion unit, determine whether it mentions any of the six Social Experience dimensions.
If a unit does not involve any Social Experience content, do not output that unit at all.
If it involves one or more dimensions, output for that unit: the matched dimension name(s) (multi-select allowed).
Notes:
If there is no subjective evaluation and only objective facts, you may choose not to score that dimension (i.e., do not include it in the dimension list), or you may assign a score of 3—but keep the style consistent.
If a sentence contains both positive and negative evaluations, decide the score based on the overall tone and the dominant tendency.
## Output Requirements & JSON Schema
Output only valid JSON. Do not include any extra text, comments, or explanations.
## Output format (example structure; field names must match exactly):
[{
    "span": "Segment text 1",
    "dimensions": ["Interpersonal Connectedness", "Brand–Self Connectedness"]
  },{
    "span": "Segment text 2",
    "dimensions": ["Community Participation"]
  }]
\end{Prompt}
}

\section{Satisfaction Sentiment Scale Guidelines (1–5)}
\label{Satisfaction Guidelines}
Table~\ref{tab:satisfaction_scale} documents the annotation guidelines for the unified satisfaction sentiment scale. We assign an ordinal satisfaction score from 1 to 5 to each extracted evidence unit, including aspect–opinion pairs for Functional and Sensory Experiences and semantic spans for Social Experiences. The score reflects the user’s evaluative attitude expressed in the evidence unit, rather than the overall stance of the full review.

\begin{table}[ht]
\caption{Satisfaction sentiment scale (1--5) used for evidence-unit sentiment annotation and modeling.}
\label{tab:satisfaction_scale}
    \centering
    \small
\begin{tabular}{|c|c|p{8.2cm}|}
\hline
\rowcolor[HTML]{C0C0C0} 
\textbf{Score} & \textbf{Label} & \textbf{Definition} \\ \hline
1 & Strong dissatisfaction & Severe negative evaluation; regret/refusal; strong disappointment (e.g., ``terrible'', ``unusable'', ``regret buying''). \\ \hline
2 & Mild dissatisfaction & Negative but limited intensity; moderate impact; minor issues dominate (e.g., ``not great'', ``a bit disappointing''). \\ \hline
3 & Neutral or mixed & Factual or balanced pros/cons; no clear stance (e.g., ``okay'', ``average'', ``not bad but\ldots''). \\ \hline
4 & Mild satisfaction & Positive with limited intensity; generally satisfied (e.g., ``pretty good'', ``works well'', ``satisfied''). \\ \hline
5 & Strong satisfaction & Strong positive evaluation; delight; strong recommendation (e.g., ``excellent'', ``love it'', ``highly recommend''). \\ \hline
\end{tabular}

\end{table}

We annotate sentiment at the evidence-unit level.
For Functional and Sensory Experiences, the evidence unit is an (aspect, opinion) pair. The sentiment score reflects the polarity and intensity of the opinion toward the aspect as expressed in the local context.
For Social Experiences, the evidence unit is a semantic span that conveys a social experience meaning. The sentiment score reflects the evaluative stance expressed toward the social experience described by that span.

\section{Detailed Performance of Branch-Specific L3/L4 Metric Classifiers}
\label{Detailed Performance}


\revise{ Tables~\ref{tab:concrete}, \ref{tab:concrete2}, and \ref{tab:concrete3} report the detailed per-class performance for the seven branch-specific L3/L4 metric classifiers used for Functional and Sensory evidence units. The results were calculated from pooled out-of-fold predictions across the five review-level grouped folds, so that each evidence unit was evaluated once as a held-out instance. For each metric, we report the number of supporting evidence units (\textit{Support}), precision (P), recall (R), and F1-score. Social Experience metric labeling was evaluated separately as SOC L3 metric classification in the LLM-based evidence-identification evaluation, because the Social Experience branch contains L3 categories but no L4 branch-specific classifier. }

\revise{ 
Because the evaluation is based on naturally occurring review evidence rather than a class-balanced benchmark, support varies across metrics. Support should therefore be interpreted together with the per-class scores. Metrics with limited support are reported for transparency and as evidence of observability in the corpus, but their scores should be interpreted descriptively rather than as stable estimates of generalization. Accordingly, the main corpus-level analyses rely on Level 2, while L3/L4 metrics are used as support-sensitive drill-down evidence. }

\revise{To further examine the reliability of fine-grained metric classification, we inspected the misclassified evidence units from the held-out folds of the seven branch-specific classifiers. Across the five review-level grouped folds, the classifiers produced 286 misclassified evidence units, corresponding to an overall error rate of 6.03\%. This indicates that the remaining errors accounted for only a small proportion of the evaluated evidence units.} 

\revise{ The remaining errors were concentrated in cases where short user-generated phrases provided limited context for distinguishing closely related fine-grained metrics. For example, expressions about interface appearance sometimes made it difficult to distinguish whether the user was referring to the lock-screen interface, screen-off display, or the broader visual design of the system. Similarly, short comments about vibration occasionally lacked enough context to distinguish vibration stability, vibration style, and vibration intensity. These cases reflect the compact and context-dependent nature of social media reviews: users often describe an experience using a brief everyday expression without explicitly specifying the evaluated object or usage condition. } 

\revise{ This error analysis suggests that the remaining misclassifications are mainly associated with ambiguous or underspecified review expressions rather than a systematic failure of the hierarchical framework. We therefore interpret L3/L4 outputs as support-sensitive drill-down evidence, while using Level 2 as the primary level for corpus-level comparison. }

\begin{table}[ht]
\caption{\revise{Detailed per-class performance of represented L3/L4 metrics (Part I: \textit{Usefulness}, \textit{Ease of Use} and \textit{Reliability}).}}
\label{tab:concrete}
 \small
\begin{tabular*}{0.9\textwidth}{@{\extracolsep\fill}lllllll}
\toprule
\revise{Metric                                                                                            } & \revise{ Level} & \revise{ Support} & \revise{ P     } & \revise{ R     } & \revise{ F1 }    \\
\midrule

\revise{Time Saving  } & \revise{ L3   } & \revise{ 175    } & \revise{ 0.9375} & \revise{ 0.9429} & \revise{ 0.9402 \\
\begin{tabular}[c]{@{}l@{}} \revise{Accuracy of   Personalized} \\ \revise{Recommendations}\end{tabular}             } & \revise{ L4   } & \revise{ 55     } & \revise{ 1.0000} & \revise{ 0.9636} & \revise{ 0.9815} \\
\revise{Application   Compatibility  }                                                                      & \revise{ L4   } & \revise{ 120    } & \revise{ 0.9160} & \revise{ 0.9083 & 0.9121 }\\
\revise{Labor Saving                                                                                      } & \revise{ L3   } & \revise{ 383    } & \revise{ 0.9427} & \revise{ 0.9452} & \revise{ 0.9439 } \\
\revise{Cross-Device   Collaboration                                                                      } & \revise{ L4   } & \revise{ 119    } & \revise{ 0.9576} & \revise{ 0.9496} & \revise{ 0.9536 } \\
\revise{Multi-Scenario   Applicability                                                                    } & \revise{ L4   } & \revise{ 184    } & \revise{ 0.8889} & \revise{ 0.8696} & \revise{ 0.8791 } \\
\revise{Adaptability   to Physical Conditions                                                             } & \revise{ L4   } & \revise{ 28     } & \revise{ 0.9630} & \revise{ 0.9286} & \revise{ 0.9455 } \\
\revise{Quick   Operations                                                                                } & \revise{ \revise{ L4   } & \revise{ 80     } & \revise{ 0.8824} & \revise{ 0.9375} & \revise{ 0.9091 } \\
\revise{Increase Usage   Benefits                                                                         } & \revise{ L3   } & \revise{ 885    } & \revise{ 0.9705} & \revise{ 0.9650} & \revise{ 0.9677 } \\
\revise{Smoothness                                                                                        } & \revise{ L4   } & \revise{ 535    } & \revise{ 0.9643} & \revise{ 0.9589} & \revise{ 0.9616 } \\
\revise{Intelligence                                                                                      } & \revise{ L4   } & \revise{ 130    } & \revise{ 0.9338} & \revise{ 0.9769} & \revise{ 0.9549 } \\
\revise{Functional   Innovativeness                                                                       } & \revise{ L4   } & \revise{ 35     } & \revise{ 0.8333} & \revise{ 0.7143} & \revise{ 0.7692 } \\
\revise{Functional   Autonomy                                                                             } & \revise{ L4   } & \revise{ 7      } & \revise{ 0.3333} & \revise{ 0.1429} & \revise{ 0.2000 } \\
\revise{Functional   Playfulness                                                                          } & \revise{ L4   } & \revise{ 54     } & \revise{ 0.8800} & \revise{ 0.8148} & \revise{ 0.8462 } \\
\revise{Functional   Richness                                                                             } & \revise{ L4   } & \revise{ 96     } & \revise{ 0.7907} & \revise{ 0.7083} & \revise{ 0.7473 } \\
\revise{Learnability                                                                                      } & \revise{ L3   } & \revise{ 55     } & \revise{ 0.9821} & \revise{ 1.0000} & \revise{ 0.9910 } \\
\revise{Learning   Threshold                                                                              } & \revise{ L4   } & \revise{ 41     } & \revise{ 0.9474} & \revise{ 0.8780} & \revise{ 0.9114 } \\
\revise{Learning Cost                                                                                     } & \revise{ L4   } & \revise{ 14     } & \revise{ 0.7500} & \revise{ 0.8571} & \revise{ 0.8000 } \\
\revise{Clarity                                                                                           } & \revise{ L3   } & \revise{ 228    } & \revise{ 0.9823} & \revise{ 0.9737} & \revise{ 0.9780 } \\
\revise{Logical   Consistency                                                                             } & \revise{ L4   } & \revise{ 140    } & \revise{ 0.9701} & \revise{ 0.9286} & \revise{ 0.9489 } \\
\revise{Layout   Rationality                                                                              } & \revise{ L4   } & \revise{ 42     } & \revise{ 0.8889} & \revise{ 0.9524} & \revise{ 0.9195 } \\
\revise{Operability                                                                                       } & \revise{ L3   } & \revise{ 75     } & \revise{ 0.9351} & \revise{ 0.9600} & \revise{ 0.9474 } \\
\revise{Task Execution   Efficiency                                                                       } & \revise{ L4   } & \revise{ 31     } & \revise{ 1.0000} & \revise{ 0.9355} & \revise{ 0.9667 } \\
\revise{Error   Tolerance                                                                                 } & \revise{ L4   } & \revise{ 23     } & \revise{ 1.0000} & \revise{ 1.0000} & \revise{ 1.0000 } \\
\revise{Memorability                                                                                      } & \revise{ L4   } & \revise{ 21     } & \revise{ 0.9524} & \revise{ 0.9524} & \revise{ 0.9524 } \\
\begin{tabular}[c]{@{}l@{}}\revise{Care for   Special Groups}  \\ \revise{(Elderly, Children, Disabled)}\end{tabular}} & \revise{ L4   } & \revise{ 46     } & \revise{ 1.0000} & \revise{ 0.9783} & \revise{ 0.9890 } \\
\revise{Stability                                                                                         } & \revise{ L3   } & \revise{ 609    } & \revise{ 0.9902} & \revise{ 0.9967} & \revise{ 0.9935 } \\
\revise{System   Stability                                                                                } & \revise{ L4   } & \revise{ 115    } & \revise{ 0.9823} & \revise{ 0.9652} & \revise{ 0.9737 } \\
\revise{Network   Stability                                                                               } & \revise{ L4   } & \revise{ 107    } & \revise{ 0.9640} & \revise{ 1.0000} & \revise{ 0.9817 } \\
\revise{Call Stability                                                                                    } & \revise{ L4   } & \revise{ 92     } & \revise{ 0.9783} & \revise{ 0.9783} & \revise{ 0.9783 } \\
\revise{Connection   Stability                                                                            } & \revise{ L4   } & \revise{ 211    } & \revise{ 0.9952} & \revise{ 0.9858} & \revise{ 0.9905 } \\
\revise{Charging   Stability                                                                              } & \revise{ L4   } & \revise{ 20     } & \revise{ 1.0000} & \revise{ 1.0000} & \revise{ 1.0000 } \\
\revise{Native   Software Stability                                                                       } & \revise{ L4   } & \revise{ 25     } & \revise{ 0.8077} & \revise{ 0.8400} & \revise{ 0.8235 } \\
\revise{Third-Party   Software Stability                                                                  } & \revise{ L4   } & \revise{ 39     } & \revise{ 0.9474} & \revise{ 0.9231} & \revise{ 0.9351 } \\
\revise{Security                                                                                          } & \revise{ L3   } & \revise{ 143    } & \revise{ 0.9786} & \revise{ 0.9580} & \revise{ 0.9682 } \\
\revise{Reduction of   Hassles                                                                            } & \revise{ L4   } & \revise{ 61     } & \revise{ 0.9508} & \revise{ 0.9508} & \revise{ 0.9508 } \\
\revise{Risk Reduction                                                                                    } & \revise{ L4   } & \revise{ 82     } & \revise{ 0.9615} & \revise{ 0.9146} & \revise{ 0.9375 } \\
\revise{Responsiveness                                                                                    } & \revise{ L3   } & \revise{ 67     } & \revise{ 1.0000} & \revise{ 1.0000} & \revise{ 1.0000 } \\
\revise{Service   Response Speed                                                                          } & \revise{ L4   } & \revise{ 29     } & \revise{ 1.0000} & \revise{ 1.0000} & \revise{ 1.0000 } \\
\revise{Service   Response Quality                                                                        } & \revise{ L4   } & \revise{ 38     } & \revise{ 1.0000} & \revise{ 1.0000 } & \revise{1.0000 } \\

\bottomrule
\end{tabular*}

\end{table}

\begin{table}[ht]
\caption{Detailed per-class performance of represented L3/L4 metrics (Part II: \textit{Durability} and \textit{Visual Appeal}).}
\label{tab:concrete2}
 \small
\begin{tabular*}{0.9\textwidth}{@{\extracolsep\fill}lllllll}
\toprule
\revise{Metric                                                                                            } & \revise{ Level} & \revise{ Support} & \revise{ P     } & \revise{ R     } & \revise{ F1     } \\
\midrule
\revise{Battery Life                                                                                      } & \revise{ L3   } & \revise{ 268    } & \revise{ 0.9963} & \revise{ 1.0000} & \revise{ 0.9981 } \\
\revise{Battery Health   Management                                                                       } & \revise{ L4   } & \revise{ 46     } & \revise{ 1.0000} & \revise{ 0.9783} & \revise{ 0.9890 } \\
\revise{Standby   Battery Life                                                                            } & \revise{ L4   } & \revise{ 48     } & \revise{ 1.0000} & \revise{ 1.0000} & \revise{ 1.0000 } \\
\revise{Gaming Battery   Life                                                                             } & \revise{ L4   } & \revise{ 47     } & \revise{ 1.0000} & \revise{ 0.9787} & \revise{ 0.9892 } \\
\revise{Daily Usage   Battery Life                                                                        } & \revise{ L4   } & \revise{ 36     } & \revise{ 1.0000} & \revise{ 1.0000} & \revise{ 1.0000 } \\
\revise{Audio-Video   Playback Battery Life                                                               } & \revise{ L4   } & \revise{ 33     } & \revise{ 0.9429} & \revise{ 1.0000} & \revise{ 0.9706 } \\
\revise{Photography   Battery Life                                                                        } & \revise{ L4   } & \revise{ 36     } & \revise{ 1.0000} & \revise{ 1.0000} & \revise{ 1.0000 } \\
\revise{Battery Life   in Power-Saving Mode                                                               } & \revise{ L4   } & \revise{ 22     } & \revise{ 1.0000} & \revise{ 1.0000} & \revise{ 1.0000 } \\
\revise{Memory   Management                                                                               } & \revise{ L3   } & \revise{ 87     } & \revise{ 1.0000} & \revise{ 1.0000} & \revise{ 1.0000 } \\
\revise{RAM Management                                                                                    } & \revise{ L4   } & \revise{ 48     } & \revise{ 0.9796} & \revise{ 1.0000} & \revise{ 0.9897 } \\
\revise{Storage   Management                                                                              } & \revise{ L4   } & \revise{ 39     } & \revise{ 1.0000} & \revise{ 0.9744} & \revise{ 0.9870 } \\
\revise{Performance   Retention                                                                           } & \revise{ L3   } & \revise{ 76     } & \revise{ 1.0000} & \revise{ 0.9868} & \revise{ 0.9934 } \\
\revise{Performance   Maintenance                                                                         } & \revise{ L4   } & \revise{ 39     } & \revise{ 1.0000} & \revise{ 0.9744} & \revise{ 0.9870 } \\
\revise{System Update                                                                                     } & \revise{ L4   } & \revise{ 37     } & \revise{ 1.0000} & \revise{ 1.0000} & \revise{ 1.0000 } \\
\revise{Appearance and   Design                                                                           } & \revise{ L3   } & \revise{ 404    } & \revise{ 0.9853} & \revise{ 0.9975} & \revise{ 0.9914 } \\
\revise{System Theme   Attractiveness                                                                     } & \revise{ L4   } & \revise{ 119    } & \revise{ 0.9262} & \revise{ 0.9496} & \revise{ 0.9378 } \\
\revise{Font   Attractiveness                                                                             } & \revise{ L4   } & \revise{ 50     } & \revise{ 0.9792} & \revise{ 0.9400} & \revise{ 0.9592 } \\
\revise{Wallpaper   Attractiveness                                                                        } & \revise{ L4   } & \revise{ 46     } & \revise{ 1.0000} & \revise{ 0.9783} & \revise{ 0.9890 } \\
\revise{Screen-Off   Attractiveness                                                                       } & \revise{ L4   } & \revise{ 26     } & \revise{ 0.6471} & \revise{ 0.8462} & \revise{ 0.7333 } \\
\revise{Home Screen   Attractiveness                                                                      } & \revise{ L4   } & \revise{ 40     } & \revise{ 0.8810} & \revise{ 0.9250} & \revise{ 0.9024 } \\
\revise{Lock Screen   Attractiveness                                                                      } & \revise{ L4   } & \revise{ 46     } & \revise{ 0.9459} & \revise{ 0.7609} & \revise{ 0.8434 } \\
\revise{Size and   Weight                                                                                 } & \revise{ L4   } & \revise{ 38     } & \revise{ 1.0000} & \revise{ 1.0000} & \revise{ 1.0000 } \\
\revise{Appearance                                                                                        } & \revise{ L4   } & \revise{ 39     } & \revise{ 1.0000} & \revise{ 1.0000} & \revise{ 1.0000 } \\
\revise{Display   Quality                                                                                 } & \revise{ L3   } & \revise{ 354    } & \revise{ 1.0000} & \revise{ 0.9746} & \revise{ 0.9871 } \\
\revise{Basic Display   Performance                                                                       } & \revise{ L4   } & \revise{ 122    } & \revise{ 0.9677} & \revise{ 0.9836} & \revise{ 0.9756 } \\
\revise{Brightness   Display Performance                                                                  } & \revise{ L4   } & \revise{ 89     } & \revise{ 1.0000} & \revise{ 1.0000} & \revise{ 1.0000 } \\
\begin{tabular}[c]{@{}l@{}}\revise{Display   Performance}\\ \revise{ under Different Modes }\end{tabular}              & \revise{ L4   } & \revise{ 143    } & \revise{ 0.9848} & \revise{ 0.9091} & \revise{ 0.9455 } \\
\revise{Image   Aesthetics                                                                                } & \revise{ L3   } & \revise{ 212    } & \revise{ 0.9636} & \revise{ 1.0000} & \revise{ 0.9815 } \\
\revise{Camera   Performance                                                                              } & \revise{ L4   } & \revise{ 40     } & \revise{ 0.9756} & \revise{ 1.0000} & \revise{ 0.9877 } \\
\revise{Video   Stability                                                                                 } & \revise{ L4   } & \revise{ 50     } & \revise{ 0.9615} & \revise{ 1.0000} & \revise{ 0.9804 } \\
\begin{tabular}[c]{@{}l@{}}\revise{Real-Time   Filters} \\ \revise{and Beautification Features}\end{tabular}          & \revise{ L4   } & \revise{ 49     } & \revise{ 1.0000} & \revise{ 1.0000} & \revise{ 1.0000 } \\
\revise{Low-Light   Photography                                                                           } & \revise{ L4   } & \revise{ 38     } & \revise{ 0.9500} & \revise{ 1.0000} & \revise{ 0.9744 } \\
\revise{Color   Reproduction                                                                              } & \revise{ L4   } & \revise{ 35     } & \revise{ 1.0000} & \revise{ 0.9714} & \revise{ 0.9855 } \\
\revise{Animation   Effects                                                                               } & \revise{ L3   } & \revise{ 55     } & \revise{ 1.0000} & \revise{ 0.9273} & \revise{ 0.9623 } \\
\revise{Animation   Naturalness                                                                           } & \revise{ L4   } & \revise{ 38     } & \revise{ 1.0000} & \revise{ 0.9474} & \revise{ 0.9730 } \\
\revise{Animation   Style                                                                                 } & \revise{ L4   } & \revise{ 17     } & \revise{ 1.0000} & \revise{ 0.8824} & \revise{ 0.9375} \\

\bottomrule
\end{tabular*}

\end{table}

\begin{table}[ht]
\caption{Detailed per-class performance of represented L3/L4 metrics (Part III: \textit{Auditory Effect} and \textit{Tactile Effect}).}
\label{tab:concrete3}
 \small
\begin{tabular*}{0.9\textwidth}{@{\extracolsep\fill}lllllll}
\toprule
\revise{Metric                                                                                            } & \revise{ Level} & \revise{ Support} & \revise{ P     } & \revise{ R     } & \revise{ F1     } \\
\midrule
\revise{Volume                                                                                            } & \revise{ L3   } & \revise{ 37     } & \revise{ 0.9375} & \revise{ 0.8108} & \revise{ 0.8696 } \\
\revise{Sound Quality                                                                                     } & \revise{ L3   } & \revise{ 38     } & \revise{ 1.0000} & \revise{ 0.9737} & \revise{ 0.9867 } \\
\revise{Call Audio   Quality                                                                              } & \revise{ L3   } & \revise{ 53     } & \revise{ 0.9423} & \revise{ 0.9245} & \revise{ 0.9333 } \\
\revise{Noise   Reduction Effect                                                                          } & \revise{ L4   } & \revise{ 14     } & \revise{ 1.0000} & \revise{ 0.8571} & \revise{ 0.9231 } \\
\revise{Call Clarity                                                                                      } & \revise{ L4   } & \revise{ 39     } & \revise{ 0.9250} & \revise{ 0.9487} & \revise{ 0.9367 } \\
\revise{Media Audio   Quality                                                                             } & \revise{ L3   } & \revise{ 134    } & \revise{ 0.9638} & \revise{ 0.9925} & \revise{ 0.9779 } \\
\revise{Video Audio   Quality                                                                             } & \revise{ L4   } & \revise{ 50     } & \revise{ 0.9615} & \revise{ 1.0000} & \revise{ 0.9804 } \\
\revise{Music Audio   Quality                                                                             } & \revise{ L4   } & \revise{ 44     } & \revise{ 0.9556} & \revise{ 0.9773} & \revise{ 0.9663 } \\
\revise{Gaming Audio   Quality                                                                            } & \revise{ L4   } & \revise{ 40     } & \revise{ 0.9268} & \revise{ 0.9500} & \revise{ 0.9383 } \\
\revise{System Audio   Quality                                                                            } & \revise{ L3   } & \revise{ 60     } & \revise{ 0.9836} & \revise{ 1.0000} & \revise{ 0.9917 } \\
\revise{Notification   Audio Quality                                                                      } & \revise{ L4   } & \revise{ 22     } & \revise{ 0.9412} & \revise{ 0.7273} & \revise{ 0.8205 } \\
\revise{Alert Audio   Quality                                                                             } & \revise{ L4   } & \revise{ 38     } & \revise{ 0.8810} & \revise{ 0.9737} & \revise{ 0.9250 } \\
\revise{Haptic   Feedback Effect                                                                          } & \revise{ L3   } & \revise{ 94     } & \revise{ 1.0000} & \revise{ 0.9894} & \revise{ 0.9947 } \\
\revise{Vibration   Stability                                                                             } & \revise{ L4   } & \revise{ 44     } & \revise{ 0.8571} & \revise{ 0.8182} & \revise{ 0.8372 } \\
\revise{Vibration   Style                                                                                 } & \revise{ L4   } & \revise{ 25     } & \revise{ 0.7083} & \revise{ 0.6800} & \revise{ 0.6939 } \\
\revise{Vibration   Intensity                                                                             } & \revise{ L4   } & \revise{ 25     } & \revise{ 0.9200} & \revise{ 0.9200} & \revise{ 0.9200 } \\
\revise{Thermal   Perception                                                                              } & \revise{ L3   } & \revise{ 124    } & \revise{ 0.9920} & \revise{ 1.0000} & \revise{ 0.9960 } \\
\revise{Charging   Temperature                                                                            } & \revise{ L4   } & \revise{ 34     } & \revise{ 0.9429} & \revise{ 0.9706} & \revise{ 0.9565 } \\
\revise{Daily Usage   Temperature                                                                         } & \revise{ L4   } & \revise{ 41     } & \revise{ 1.0000} & \revise{ 0.9512} & \revise{ 0.9750 } \\
\revise{Gaming   Temperature                                                                              } & \revise{ L4   } & \revise{ 49     } & \revise{ 0.9412} & \revise{ 0.9796} & \revise{ 0.9600 } \\
\revise{Touch Response                                                                                    } & \revise{ L3   } & \revise{ 102    } & \revise{ 1.0000} & \revise{ 1.0000} & \revise{ 1.0000 } \\
\revise{Touch   Stability                                                                                 } & \revise{ L4   } & \revise{ 48     } & \revise{ 0.9796} & \revise{ 1.0000} & \revise{ 0.9897 } \\
\revise{Touch Accuracy                                                                                    } & \revise{ L4   } & \revise{ 42     } & \revise{ 0.9767} & \revise{ 1.0000} & \revise{ 0.9882 } \\
\begin{tabular}[c]{@{}l@{}}\revise{Touch   Performance}  \\\revise{ under Special Conditions }\end{tabular}            & \revise{ L4   } & \revise{ 12     } & \revise{ 1.0000} & \revise{ 0.8333} & \revise{ 0.9091 } \\
\revise{Surface   Texture                                                                                 } & \revise{ L3   } & \revise{ 26     } & \revise{ 1.0000} & \revise{ 1.0000} & \revise{ 1.0000} \\
\bottomrule
\end{tabular*}

\end{table}


\end{document}